# Exoplanet Science From *Kepler*

**Jack J. Lissauer**
*Space Science & Astrobiology Division, MS 245-3, NASA Ames Research Center, Moffett Field, CA 94035, USA*

**Natalie M. Batalha**
*Department of Astronomy and Astrophysics, University of California, Santa Cruz, CA 95060, USA*

**William J. Borucki**
*Astrophysics Branch, MS 244-30, NASA Ames Research Center, Moffett Field, CA 94035, USA*

The *Kepler* spacecraft, whose single instrument was a 0.95 m diameter wide-field telescope, operated in a heliocentric orbit for nearly a decade, returning a wealth of data that have revolutionized exoplanet science. *Kepler* data have been used to discover thousands of planets, including hundreds of multi-planet systems. *Kepler* discoveries have greatly expanded the diversity of known exoplanets and planetary system properties. Moreover, *Kepler* has provided the best estimates of exoplanet occurrence rates as functions of planetary sizes, orbital periods and stellar type, with precise values for planets with $P \lesssim 1$ yr. We provide herein an overview of the mission and its major findings regarding the occurrence rates of planets, the mass-radius relationship for exoplanets and the architectures of planetary systems.

## 1. INTRODUCTION

In March of 2009, NASA launched the *Kepler* spacecraft into an Earth-trailing heliocentric orbit. The goal of the *Kepler* Mission was the exploration of the abundance and characteristics of planetary systems, with a special emphasis on the occurrence rates of Earth-size planets in and near the habitable zone of their stars (Sec. 3.1 of *Borucki* 2016). *Kepler* is a 0.95 m aperture space telescope with a wide field-of-view (*Borucki et al.* 2010; *Koch et al.* 2010). Above Earth's atmosphere, and away from the glare and thermal variations of low Earth orbit, *Kepler* monitored the brightness of many thousands of stars at 30 minute cadence for 4 years, in search of periodic dips caused by planetary transits (partial eclipses) of the stellar disks. High precision photometric measurements are required because while a transit of a Jupiter-size planet across the disk of a Sun-like star blocks 1% of the stellar flux and can easily be identified using a ground-based telescope, an Earth-size transiting planet fractionally reduces the flux of such a star by only $8 \times 10^{-5}$ (80 parts per million). In its 4 year prime mission, which ended in May 2013 due to the failure of a second reaction wheel that made precise stable pointing at *Kepler*'s star field impossible, *Kepler* observed over 200,000 stars, including more than 175,000 main sequence dwarfs.

Signatures of more than 4350 candidate exoplanets have been identified in *Kepler* data (Figure 1). Thorough analysis of the *Kepler* data and the ancillary observations indicate that the vast majority of the current list of planet candidates found by the *Kepler* Mission represent true exoplanets (*Morton and Johnson* 2011; *Fressin et al.* 2013; *Rowe et al.* 2014; *Bryson et al.* 2021). These worlds have greatly expanded the zoo of known exoplanet types. Most have sizes intermediate between those of Earth and Neptune, a regime not observed within our Solar System. Approximately 40% are in systems where multiple transiting planets have been found, and *Kepler* data are telling us that flat systems, some containing multiple planets on closely-spaced orbits, are quite common. Most *Kepler* planets have orbital periods are shorter than a few months, although this is largely a consequence of observational biases of the transit method, compounded by the limited duration of the survey.

The history of many *Kepler* exoplanet firsts is covered by a collection of articles contained in a virtual special issue of the journal *New Astronomy Reviews* (*Lissauer and Eisberg* 2018; *Ragozzine and Holman* 2018; *Torres and Fressin* 2018; *Agol and Carter* 2018; *Borucki et al.* 2018; *Winn et al.* 2018; *Steffen and Lissauer* 2018; *Nesvorný* 2019; *Doyle* 2019).

Many individual *Kepler* planets and planetary systems were exciting and illuminating discoveries. But *Kepler* is, in essence, a statistical mission, designed to determine the abundance of planets down to the size of Earth orbiting within and interior to circumstellar habitable zones (HZs), where planets with an atmosphere similar to Earth's receive the right amount of stellar radiation to maintain reservoirs of liquid water on their surfaces.

We provide an overview of the *Kepler* Mission in Sec. 2. We discuss the differences between planetary candidates and confirmed planets in Sec. 3. We present some of the most interesting planets and planetary systems found by *Kepler* in Sec. 4. Section 5 summarizes findings of some of the key studies of planetary occurrence rates as functions of radius and orbital period using *Kepler* data. The majority of small and mid-sized exoplanets with both radius and mass measurements prior to 2020 were *Kepler* discoveries; we present some of the key results and their implications for planetary compositions in Sec. 6. The precision of *Kepler* photometry has enabled the characterization of some short-period planets (primarily hot jupiters) from phase variations observed outside of the planetary transits (Sec. 7). We summarize key *Kepler* results regarding the architectures of planetary systems in Sec. 8. Section 9 concludes this chapter with a brief summary and a discussion of future exoplanet transit studies.

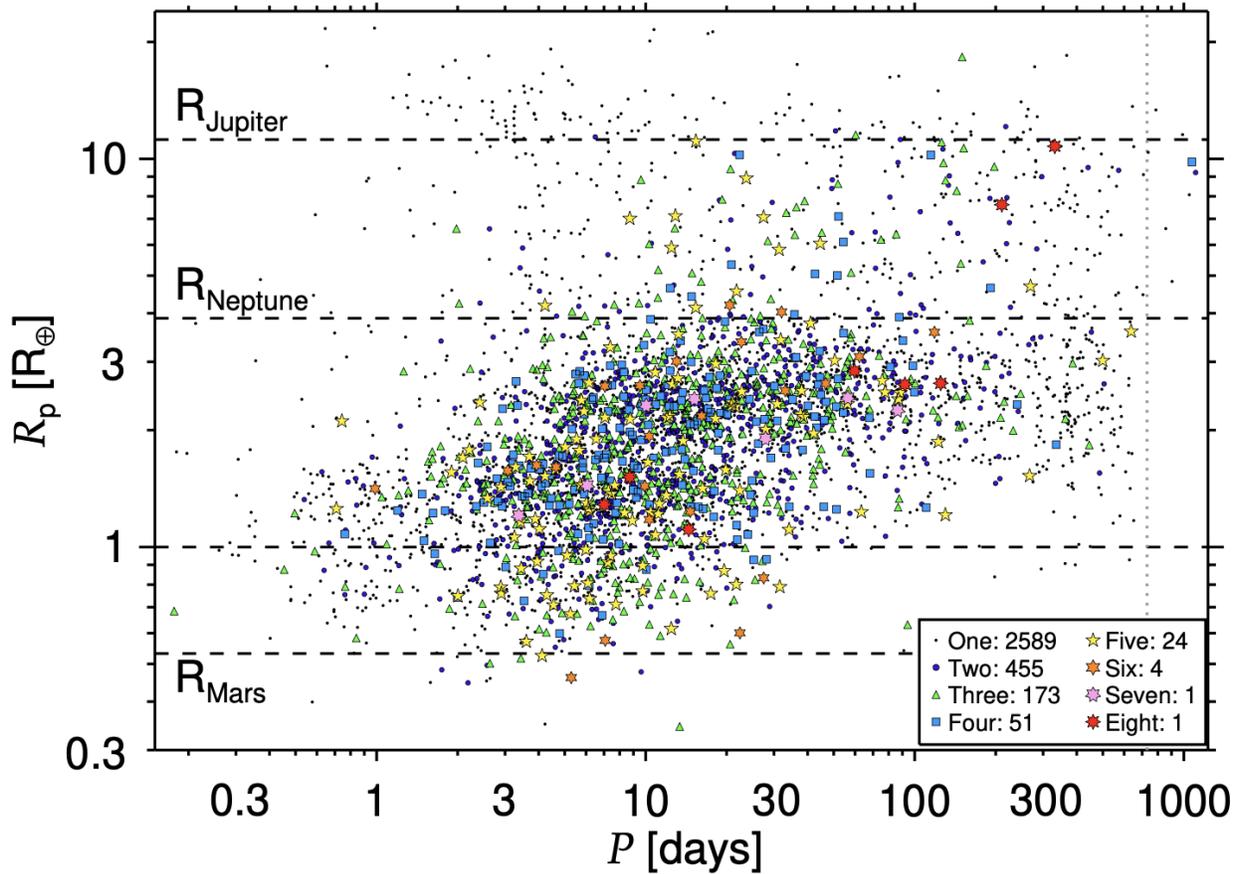

Fig. 1.— Radii and orbital periods of *Kepler*'s planetary candidates. Those planets that are the only candidate for their given star are represented by black dots, those in two-planet systems as dark blue circles, those in three-planet systems as green triangles, those in four planet systems as light blue squares, five candidates as yellow five-pointed stars, six planets as orange six-pointed stars, the seven planet candidates associated with KOI-2433 and the eight planets orbiting KOI-351 (Kepler-90) as red eight-pointed stars (from the revised (R1) version of *Lissauer et al.* 2023).

## 2. *Kepler* MISSION OVERVIEW

*Kepler* was selected as NASA's tenth Discovery Mission in December of 2001. The spacecraft was launched in March 2009 and conducted its prime scientific mission observing a single star field with a duty cycle of ∼ 90% from May 2009 – May 2013. A second mission, referred to as *K2*, limited by the capabilities of an aging spacecraft and reduced budget, observed twenty (mostly non-overlapping) star fields for short periods of time from March 2014 – September 2018. The development of the concept that ultimately became *Kepler*, the history of proposals to NASA and the organizational structure of the project are reviewed in *Borucki et al.* (2003) and *Borucki* (2016).

### 2.1. Spacecraft Characteristics, Capabilities and Performance

The *Kepler* spacecraft contained only one instrument: the telescope/photometer (*Koch et al.* 2010). The photometer can be considered to be a very large camera that orbited the Sun and almost continuously recorded images from a single region of the sky. The 0.95 m diameter Schmidt-type telescope (Sec. 2.3.4 of Troeltzsch & Howell 2020) was chosen to provide a sufficient number of photons from bright (i.e., $12^{\text{th}}$ magnitude) stars to reliably recognize three or more transits of Earth-size planets across Sun-like stars. Its field of view covered 113 square degrees (*Borucki* 2016). Data storage, downlink capability and computer software allowed simultaneous time-series measurements of the brightness of up to 170,000 selected stars (*Batalha*

*et al.* 2010b).

The spacecraft was launched into a 372.5-day heliocentric orbit to provide a stable thermal environment and nearly continuous observations 24 hours per day throughout the orbit. Breaks in the photometric time series data occurred when the spacecraft was rotated on its optical axis every quarter-orbit (i.e., ∼ 93 days) to keep the solar panels pointed at the Sun and the sunshade properly positioned to prevent sunlight from entering the telescope (*Koch et al.* 2010). Quarters were numbered consecutively, beginning with the start of regular science operations in Quarter 1 (Q1) through the end of *Kepler*'s Prime Mission in Q17. About 50,000 targets were observed for 10 days primarily as an engineering test during the commissioning phase of the mission; these data have also been studied scientifically, and this time interval is referred to as Q0. A single detector module (Module #3; there were a total of 21 science modules) failed on 9 January 2010 (*Borucki* 2016); consequently, the four regions of the sky that it observed were monitored only three-fourths of each year for the remainder of the prime mission.

Shorter breaks in the photometric time series data occurred on a monthly basis, when the spacecraft was re-oriented to point its high-gain antenna toward the Earth to downlink the data. The downlinked data were forwarded to NASA Ames Research Center, where they were analyzed, searched for transit events, and checked for false positives. Both the raw and processed data and light curves were released to the public and archived at the Mikulski Archive for Space Telescopes at the Space Telescope Science Institute. Planet candidates and other transit-like signals identified in the light curves were archived at the NASA Exoplanet Archive at the NASA Exoplanet Science Institute.

To maximize the observing time and to minimize the number of missed transits, a single star field was observed during the prime operational phase of the *Kepler* Mission, prior to the *K2* Mission phase. To prevent scattering of sunlight into the telescope, the telescope was designed to have a 55° Sun-avoidance angle. To maximize the number of uncrowded dwarf-type stars in the field of view (FOV), the field was chosen to be slightly off the galactic plane, centered at R.A. = 19h22m40s and declination = $43°30'00''$ (*Koch et al.* 2010).

Images of the single star field near the Cygnus/Lyra constellations were taken every 6 seconds. Various techniques were used to reduce the onboard data storage and download requirements: Data from only a small number of pixels for each of the selected stars (typically ∼ 20, but depending on the brightness of the target) were down-linked to the ground. Downlink requirements were further reduced by a data compression protocol (*Jenkins and Dunnuck* 2011) and by stacking the star images to provide 30-minute (more precisely 29.4 minutes) "long cadence" sums for the pixels covering each of the ∼ 170,000 selected stars. Full-frame images of the field-of-view (long-cadence duration) were also downloaded roughly once per month, primarily to confirm the proper orientation and placement of the detectors on the sky and to assess photometer health.

Additionally, each Quarter the brightness values of 512 of the target stars were also downloaded at a "short cadence" of 1 minute. Some short cadence targets were on the list for most of the mission, whereas others were only selected for one or two quarters. This short cadence was chosen to better define the duration and epochs needed to measure transit-time-variations of planetary transits (*Steffen* 2016), to provide estimates of planetary orbit eccentricity (*Moorhead et al.* 2011; *Lissauer et al.* 2023), and to enable asteroseismic determination of stellar size and age (*Chaplin et al.* 2014). Period analysis of the photometric variations of stellar brightness show a rich spectra of overtones of oscillations and pulsations from pressure mode (p-mode) and gravity-mode (g-mode) waves for a variety of star types including dwarfs, subgiants, and red giants. These measurements allow model-based estimation of fundamental stellar properties, including size, density, and age. For many of the bright planet-hosting main-sequence stars (i.e., *Kepler* magnitude (Kp) > 11 and $T_{\rm eff}$ > 5300 K), the sizes and densities derived from the asteroseismic investigations provide accurate values (∼ 2%) that allow the characteristics of both the planets and the stars they orbit to be deduced (*Lundkvist et al.* 2018).

Evaluations of 6.5 hour integrations (typical of transit durations) showed a photometric precision of ∼ 6 ppm for bright, very quiet early-type stars and 29 ppm over 30 minute intervals for typical $12^{\rm th}$ magnitude G dwarfs (*Gilliland et al.* 2011). This very high photometric precision was made possible by the low pointing jitter (∼ $0.001''$) that was provided by the guidance system (*Gilliland et al.* 2011). The several sources of instrument noise that were encountered are described in *Caldwell et al.* (2010).

Pixel data downloaded from the spacecraft were converted to instrumental fluxes via *Kepler* pipeline software modules that calibrated pixel data (*Quintana et al.* 2010), performed aperture photometry (*Twicken et al.* 2010) and corrected for systematic errors (*Stumpe et al.* 2012; *Smith et al.* 2012; *Stumpe et al.* 2014). The pipeline software is documented in the *Kepler Data Processing Handbook* (KSCI-19081) at MAST. See §1.2 of *Tenenbaum et al.* (2012) for a succinct description of the *Kepler* Data Analysis Pipeline.

After the data were calibrated and systematic errors removed (*Tenenbaum et al.* 2010; *Stumpe et al.* 2014), search algorithms were employed to find the periodic dimming of stars that represent planetary transits (*Jenkins et al.* 2010). At that point, transits, statistical fluctuations of the flux measurements, and other astrophysical events that mimic transits, became evident (*Brown* 2003). Several methods were used to distinguish between valid planet candidates and these noise sources (*Bryson et al.* 2013). For example, false-positive events due to the presence of nearby variable stars could often be recognized by noting the movement of the brightness centroid of the stellar image that occurred during the transit. Doubly-eclipsing binary stars too close to the target star to be resolved by astronomical telescopes could sometimes be identified by comparing the depths of the odd and even "transits". See the *Kepler Data Processing Handbook* (KSCI-19081-002) for a comprehensive discussion of the pipeline and data processing.

To get an accurate estimate of the occurrence rate of exoplanets, it was necessary to further reduce the number of false-positive events and learn how to estimate the remaining false positive rate from *Kepler* data alone. Data from both ground-based and space-based telescopes other than *Kepler* were therefore used to confirm numerous candidates as planets and to identify others as false-positives. Many different ground-based and space-based telescope campaigns were mounted to separate valid candidates from false positives and to characterize the stars with planets (*Batalha* 2014). For example, the *Spitzer Telescope* was used to detect the color change that occurs when the dimming is caused by the occultation of a background star rather than by a planetary transit (*Désert et al.* 2015). Observations of the Doppler shift in the spectrum of the target star by very-high-resolution spectrometers at major telescope facilities were especially helpful in making mass determinations to separate orbiting planets from low mass stars (*Latham et al.* 2010; *Santerne et al.* 2012; *Marcy et al.* 2014). Speckle measurements in the near-IR and high-resolution images in the visible were employed to identify field stars that were within the pixels used to measure the brightness of the target stars (*Howell et al.* 2021). The flux from field stars in the target aperture dilutes the flux from the target star, thereby reducing the estimated size of the transiting object. The reduction was sometimes so severe that eclipsing binary stars might be mistaken as a star with a transiting planet.

The flux change that occurs during a transit only provides an estimate of the ratio of the area of the planet to the area of the star it orbits. Therefore, the estimated size of the planet and its uncertainty are directly dependent upon the knowledge of the size of the star. Without detailed examination by ground-based telescopes, stellar sizes often added 30% to the uncertainties introduced by shot noise (*Berger et al.* 2018). Recently, the measurements of the spectra of many stellar hosts of exoplanet candidates (*Petigura et al.* 2017) and the astrometric and photometric observations of most the *Kepler* target stars by the *Gaia* satellite have dramatically improved estimates of stellar size and thereby the sizes of exoplanets (*Berger et al.* 2020a; *Fulton and Petigura* 2018).

**2.2. Target Selection for *Kepler*'s Prime Mission**

A primary goal of the *Kepler* Mission was to "detect Earth-size planets in the habitable zone of solar-like stars (F to K dwarfs), determine their frequency, and identify their characteristics" (*Koch et al.* 2010; *Borucki* 2016); here the word "frequency" refers to what is now known as "occurrence rate". Neither a magnitude-limited nor volume-limited survey would accomplish that goal. The former would generate a target list favoring evolved stars. The latter would favor M dwarfs and would not have been possible for a deep survey due to catalog incompleteness. Balance was achieved by designing a selection function based on analytic detectability metrics (*Batalha et al.* 2010b).

Stellar properties like apparent magnitude, radius, photometric crowding, and extinction are necessary inputs to planet detectability estimates. A pre-launch effort to characterize stars in the FOV was initiated in 2004 using broad and narrow band photometry from existing catalogs (2MASS JHK) and observations using KeplerCam (g,r,i,z,D51) on the Whipple Telescope. The resulting *Kepler* Input Catalog (KIC) provides fundamental stellar properties for over 1 million stars in the FOV and is over 90% complete for stars brighter than $16^{\text{th}}$ magnitude in the optical (*Brown et al.* 2011). Restricting target selection to stars brighter than $16^{\text{th}}$ magnitude resulted in a dearth of late-type M dwarfs (cooler than 3000 K). It should be noted, however, that ancillary catalogs (e.g., proper motion, Gliese, Hipparcos) were utilized to add targets missed by the selection function. Bright stars lacking stellar classification in the KIC were also observed at the outset and retained or dropped based on their photometric characteristics.

To ensure long baseline photometry for a sufficiently large sample, the bulk of the target list was "locked in" at the time of launch, although small variations in the long-cadence target list and more substantial changes of the short-cadence targets were made from quarter to quarter.

By November 2009, the *Kepler* target selection group had improved methods for calculating the crowding metric and expected S/N (signal to noise ratio) of transiting planets of a given size and period around stars in the *Kepler* field. This resulted in a reprioritization that disfavored $\sim 16\,500$ stars (mostly evolved) in favor of fainter main sequence stars. None of these stars had prior KOI identifications, though about 100 had prior TCEs. Approximately 5000 of the stars that were disfavored, including all of the ones with TCEs, were ultimately retained. These changes, affecting less than 10% of the target list, account for the relatively high number of new targets observed beginning in Quarter 4.

Module 3 of *Kepler*'s focal plane, comprising two of *Kepler*'s 42 science CCDs, failed in early 2010, which restricted observations of almost 20% of the targets to three quarters per year for the remainder of the mission. The space this freed up for data storage was used to add several thousand new targets that were just below the cutoff for inclusion in the initial target list.

A few thousand independently identified targets were added by the *Kepler* Asteroseismic Consortium and Guest Observer (GO) Program on a quarter by quarter basis. The lightcurves of all target stars were searched for transits. When a planet candidate was identified around a target that had been or was scheduled to be removed from the target list, said star was observed for the remainder of the mission.

The project strived for F to K dwarf completeness in the face of uncertain or lacking information knowing that the state of knowledge would improve over the following decades, through ground-based observing campaigns, asteroseismic analyses, and the European Space Agency's *Gaia* Mission. The *Kepler* Input Catalog was updated with improved star properties after observations of *Kepler*'s prime FOV had been completed but prior to mission close-out (*Huber et al.* 2014; *Mathur et al.* 2017). *Gaia*'s second data release (DR2) (*Gaia Collaboration et al.* 2018; *Lindegren et al.* 2018) enabled parallax-based stellar luminosity and radius determinations as well as isochrone-based stellar ages (*Berger et al.* 2018, 2020a) for $\sim 90\%$ of the targets observed by *Kepler*. Moreover, the catalog enabled the characterization of the stars in the FOV that *Kepler* did not observe for a post-mission lookback on target selection performance. *Kepler* observed $\sim 206,000$ unique stars over the course of its prime mission, approximately $\sim 80,000$ of which are brighter than $14^{\text{th}}$ magnitude in the *Kepler* bandpass. That sample is over 90% complete for G and K dwarfs as identified by *Gaia* parallaxes (*Wolniewicz et al.* 2021). Completeness drops to $\sim 80\%$ and $\sim 60\%$ considering stars brighter than $15^{\text{th}}$ and $16^{\text{th}}$ magnitude, respectively. Fainter stars are more impacted by photometric contamination from nearby stars, referred to as "crowding" – a consideration that is quantified and incorporated into the planet detectability metric. Some of the degradation in completeness is, therefore, by design, but it adds additional complexity to demographic studies.

### 2.3. The *K2* Mission

The *Kepler* spacecraft observed the same field of view, and for the most part the same target stars, for 4 years during its primary mission. The original field of view targeted by *Kepler* was well above the plane of the spacecraft's orbit around the Sun (which is nearly identical to Earth's orbital plane, the ecliptic), so that observations could be made year-round. *Kepler* required at least three of its four gyroscopic reaction wheels to point stably, and when two wheels failed (wheel #2 on 16 July 2012 and wheel #4 on 14 May 2013; *Borucki* 2016), it could not continue to observe this northerly FOV.

However, the two still-functional reaction wheels were sufficient to point the telescope within the plane of *Kepler*'s orbit, because the net torque on the spacecraft from solar radiation pressure is small for this orientation. Therefore, the spacecraft was re-purposed to make $\sim 80$ day long observations of fields near the ecliptic plane, and its new mission was named *K2*. Two more CCD modules failed after the completion of the prime mission, Module 7 during an engineering test on 21 January 2014 prior to the start of the *K2* Mission and Module 4 on 20 July 2017. The remaining 18 modules continued to function until the supply of gas used to correct the pointing of the telescope was exhausted, terminating the *K2* phase of the mission, and the *Kepler* spacecraft was "retired" in October 2018, after almost a decade of service.

Target selection for *K2* was community-driven. Both the exoplanet and astrophysics communities had input on the pointing of the telescope, within the limits allowed by two-wheel operations. Although the main science objective of *K2* was to search for transiting exoplanets, other astrophysical objectives were a factor in some of the spacecraft pointings as well as for target selection.

With much shorter observations of individual targets, fewer targets observed at the same time (because more pixels were downloaded per target to partially compensate for the poorer pointing of the telescope) and somewhat poorer photometric precision because only two reaction wheels were available to stabilize pointing, *K2* could not match *Kepler* either in terms of finding small and/or long-period planets or assessing planetary occurrence rates. However, over the course of more than four years, *K2* observed an order of magnitude more area on the sky. Therefore it was able to search for short-period planets around more of the best targets for follow-up observations by other telescopes, bright Sun-like stars and moderately bright small stars.

### 3. PLANET CANDIDATES vs. VERIFIED PLANETS

The vast majority of "known" exoplanets have been detected by just a single observational technique. The first claimed detections of extrasolar planets were subsequently refuted, withdrawn, or simply never confirmed despite further observations with superior telescopes. Thus, when the first real exoplanets were detected in the latter part of the $20^{\text{th}}$ century and the beginning of the $21^{\text{st}}$ century, a great deal of evidence was required for most astronomers to accept the findings as verified discoveries. In most cases, the arguments were based on demonstrating the veracity and quality of the data and rejection of other possible interpretations, or at least showing that they were exceedingly unlikely. For a small fraction of the data, affirmations of a different sort have been obtained, allowing a confirmation of sorts. This could be detection of the signature of mutual gravitational perturbations on the orbits of planets in a multi-planet system or by a completely independent method of observation. The first example of the latter, and also the first transiting planet known, was HD 209458 b, which was discovered using the radial velocity (RV) method and subsequently observed to transit its star (*Henry et al.* 1999; *Charbonneau et al.* 2000).

For each exoplanet detection technique, early findings were controversial. As observations improved and confidence in specific detection methods increased, detection via a single method became more accepted, although various types of analysis remained requirements for a planet to be accepted. For example, radial velocity signatures, especially short-period sinusoidal ones, needed to be compared with contemporaneous indicators of stellar activity to demonstrate that the observed variations were not intrinsic to the star.

The first photometric surveys for transiting exoplanets, conducted from the ground, were plagued with false posi-tives, the majority of which were eclipsing binary stars. By the time *Kepler* was launched in 2009, improved observations and analysis techniques had reduced the fraction of false positives among candidate transiting planets found by the leading search teams. Nonetheless, the only transiting planet candidates to be generally accepted as true exoplanets were those that had been confirmed by RV observations.

Transit patterns in *Kepler* lightcurves were identified using automated searches for periodic sequences of three or more transit-shaped dips of similar duration and depth, as well as through inspections by eye, with the latter method not as sensitive to shallow transits but not restricted to planets that transited at least three times during the *Kepler* observations. Lightcurves with such transit-like characteristics were cataloged, with the target stars given *Kepler* Object of Interest (KOI) numbers (integer portion), with patterns for a given target denoted by numbers after the decimal point starting with .01. For example, KOI-157.06 referred to the sixth such pattern associated with the star denoted KOI-157. KOI's were vetted by a portion of the *Kepler* team, with the ones consistent with planetary transits classified as planet candidates (*Batalha et al.* 2010a). Those planet candidates that were subsequently verified as true exoplanets were assigned Kepler names, with, e.g., KOI-157.06 becoming Kepler-11 b.

Planets in multiple planet systems perturb one another's orbits, causing paths to deviate from purely keplerian ellipses and orbital frequencies to differ from strict periodicity (*Newton* 1687), thereby leading to changes in the intervals between transits, i.e., transit timing variations (TTVs; *Holman and Murray* 2005; *Steffen and Agol* 2005). The presence and characteristics of these TTVs provide a quasi-independent (from the shape and repetition of the transits) piece of evidence that confirm the signatures observed in a *Kepler* lightcurve are indeed caused by planets and that these planets orbit the same star. As of 2013, the majority of (the then dozens of) verified *Kepler* planets had been confirmed via analysis of TTVs (e.g., *Steffen et al.* 2012b; *Ford et al.* 2012; *Fabrycky et al.* 2012; *Steffen et al.* 2013; *Xie* 2013). But TTV confirmation was only possible for a small minority of the several thousand *Kepler* planet candidates, so other techniques were developed to verify hundreds and ultimately thousands of *Kepler* planet candidates as high-reliability known exoplanets.

The first *Kepler* planet to be validated by demonstrating that the planetary hypothesis was overwhelmingly more likely than any other interpretation of the pattern of periodic dips observed in the *Kepler* lightcurve was Kepler-9 d (*Torres et al.* 2011). Kepler-9 d is a small inner companion to two near-resonant giant planets that had already been confirmed using TTVs (see Sec. 4). The validation was accomplished using the `Blender` procedure to evaluate the relative likelihoods of planets vs. eclipsing binary stars using information from the *Kepler* lightcurve, pixel-level *Kepler* data to determine the celestial coordinates of the transit-like signature, and various ground-based data. Although `Blender` requires too much effort to be used for large numbers of planet validations, it established the principle of verifying *Kepler* candidates without confirmation by affirmative evidence of their presence other than their periodic transit lightcurve, and it has been used to verify some of *Kepler*'s most important planet candidates, such as the potentially-habitable Kepler-62 f (Sec. 4; *Borucki et al.* 2013, 2018).

The first batch-validation of hundreds of *Kepler* planet candidates was completed in early 2014, using a combination of careful lightcurve analysis together with the demonstrated substantial increase in reliability of planet candidates when two or more candidates are associated with the same *Kepler* target (*Lissauer et al.* 2012, 2014b; *Rowe et al.* 2014). The aforementioned study added more than 700 new planets with reliability substantially above 99%, thereby more than tripling the number of verified *Kepler* planets and increasing the number of *Kepler* exoplanets to more than that of all other exoplanet discoveries. Two years later, validation using a simplified technique analogous to the `Blender` procedure added more than 1000 "verified" planets to the *Kepler* list (*Morton et al.* 2016); although the reliability of this group of claimed validations was not as high as that of the 2014 study that used multiplicity as one of its requirements for validation, $\sim 99\%$ of them appear to be *bona fide* planets.

There are now more verified (confirmed or validated) *Kepler* planets as there are unverified *Kepler* planet candidates, and the former list undoubtedly has a much smaller rate of false positives as the latter. Nonetheless, *because the process of planet verification is far more biased in terms of planet and planetary system properties than is the sample of planet candidates, planet candidate lists, rather than confirmed planet lists, are more appropriate to use for statistical studies of the population of exoplanets, including studies of planetary occurrence rates*. Improved reliability of planet candidate lists can be achieved by removing candidates with low signal-to-noise ratios and/or low vetting scores.

## 4. A STROLL THROUGH *Kepler*'S EXOPLANETARY ZOO

Perhaps the most surprising finding of the first three decades of exoplanetary science is the amazing diversity among planets and planetary systems, and *Kepler* revealed more of this diversity than any other telescope. In this section, we present some of the highlights. As with any visit to a large zoo, there is far more to see than one has time to look at, displays near the entrance tend to receive extra attention, one lingers at the more exotic species, some animals are easier to observe than others, and individual tastes vary. Also, some animals are of great interest as individuals, whereas the interactions between individuals are the most exciting aspects of other exhibits.

*Kepler*'s first major exoplanet discovery was the Kepler-9 planetary system, which includes two transiting giant planets having orbital periods, $P$, of 19.2 and 38.9 days.

The planets' proximity to the 2:1 mean motion resonance (MMR) causes their mutual gravitational perturbations to induce large TTVs in Kepler-9's giant planets. Analysis of the first seven months of these TTVs enabled the planets to be confirmed and provided rough estimates of their masses (*Holman et al.* 2010). Studies using the entire four years of *Kepler* observations plus follow-up transit observations from the ground led to much more precise estimates of the planetary masses and orbits (*Freudenthal et al.* 2018). Both of these planets are Saturn-sized but less than half as massive as Saturn (*Ragozzine and Holman* 2018). As noted in Sec. 3, Kepler-9 also has a small transiting planet of unknown mass with an orbital period $P = 1.6$ days. Kepler-18 is analogous to Kepler-9, with two Neptune-mass planets near the 2:1 orbital resonance and a smaller inner planet (*Cochran et al.* 2011). Although this type of system is represented by two of *Kepler*'s early discoveries, similar systems are uncommon, with the two early discoveries resulting from the novelty of planetary systems with these characteristics combined with the relative ease by which they could be confirmed (see Sec. 3).

The first rocky planet found by *Kepler* was Kepler-10 b (*Batalha et al.* 2011), which has a mass of $M_p = 4.6$ M$_\oplus$, a radius of $R_p = 1.4$ R$_\oplus$ and an orbital period of only $P = 20$ hours. Kepler-10 b's high density results from compression caused by high internal pressure, and its bulk properties are consistent with an Earth-like composition. This planet is so strongly illuminated that, despite its small size, its phase variations and occultation by the star are evident in phase-folded *Kepler* lightcurves. The sub-stellar region on Kepler-10 b's surface is hotter than the temperature at which silicates melt, so it has been referred to as a lava world. Even shorter period planets have subsequently been found by *Kepler*. The well-studied Kepler-78 b has an extremely short period, orbiting its star in just 8.5 hours (*Sanchis-Ojeda et al.* 2013). This scorching planet is slightly larger than Earth, and its mass, measured from the radial velocity variations it induces in its nearby host star, implies a rocky composition. An analysis of *Kepler* data shows that $\sim 0.5\%$ of G dwarf stars have a planet in the size range $0.8 - 2$ R$_\oplus$ with period shorter than 1 day; such ultra-short period planets (or USPs) seem to be more common about cooler dwarf stars (*Sanchis-Ojeda et al.* 2014). Very few planets with radii $> 2$ R$_\oplus$ and $P < 1$ day have been found, despite the ease in detecting such planets via transit photometry.

Kepler-11 is a sunlike star with six transiting planets whose sizes range from $\sim 1.8 - 4.2$ R$_\oplus$. Five of these planets have orbital periods between 10 and 47 days, implying a very closely-packed dynamical system; the outer planet, Kepler-11 g, has $P = 118.4$ days. The planets' masses have been measured using TTVs. Most if not all have a substantial fraction of their volume occupied by the light gases H$_2$ and He. Observations of low density sub-neptune ($M_p < $ M$_{\rm Neptune}$) exoplanets such as those in the Kepler-11 system imply that H/He can dominate the volume of a planet that is only a few times as massive as the Earth. Although few planetary systems share Kepler-11's dynamical architecture, warm gas-rich sub-neptune planets, absent in our Solar System and first clearly characterized in the Kepler-11 system (*Lissauer et al.* 2011a, 2013), have proven to be quite common (Fig. 1).

The first transiting circumbinary planet to be identified, Kepler-16 b, is a Saturn mass, Saturn size object traveling on a nearly circular orbit around an eclipsing pair of stars, one of which is two-thirds the size of our Sun and the other only a fifth as large as the Sun. More than ten other circumbinary planets have been found using *Kepler* data. Three planets, all transiting, are known to orbit the Kepler-47 binary, which is the only known circumbinary multi-planet system (*Orosz et al.* 2019).

Kepler-20 e was the first planet smaller than Earth to be verified around a main sequence star other than the Sun (*Fressin et al.* 2012); its 6-day orbit means that at least its lit hemisphere is far too hot to be habitable. Kepler-37 b, only slightly larger than Earth's Moon, was the first planet smaller than Mercury to be found orbiting a normal star; its orbital period is 13 days, and the stellar host is 80% as massive as the Sun (*Barclay et al.* 2013). *Kepler* has found hundreds of planets/planet candidates similar in size to Earth, with orbital periods ranging from just over 4 hours to more than 100 days.

Kepler-36 hosts two planets, both significantly larger than Earth, whose semimajor axes differ by little more than 10% but whose compositions are dramatically different (*Carter et al.* 2012). The proximity of these planets' orbits leads to large TTVs, and together with astero-seismic analysis of their bright host star's oscillation using *Kepler* short-cadence lightcurves, allows for very precise characterization of the planets' bulk properties. The inner, rocky, Kepler-36 b has a radius of 1.5 R$_\oplus$ and mass of $\approx 4.4$ M$_\oplus$, consistent with a mixture of silicates and iron in the proportion that they are present in the bulk Earth. In contrast, low-density Kepler-36 c has a radius of 3.7 R$_\oplus$ and mass of $\approx 8$ M$_\oplus$, implying that most of its volume is filled with H/He. No other pair of known planets are as proximate and strongly-interacting.

Kepler-51 hosts three extremely low-density planets. These three planets, whose orbital periods range from 30 – 130 days, are all substantially larger than Neptune, yet have masses only a few times that of Earth (*Masuda* 2014; *Jontof-Hutter et al.* 2021). The three Kepler-51 planets were the first members of a class of planets now known as super-puffs to be identified, and they remain the prototypes of this uncommon and unpredicted class of planets, although additional examples such as Kepler-79 d (*Jontof-Hutter et al.* 2014) have subsequently been found. Initial models of these planets suggested that the super-puffs needed to be $\gtrsim 50\%$ H/He by mass, which would be very difficult to form according to current planetary accretion models (*Bodenheimer et al.* 2018). However, recent work by *Gao and Zhang* (2020) suggests that photochemical hazes in gas escaping from planets could be opaque at sub-microbar pressure levels, allowing the large apparent sizes of the planets to be explained with planetary bulk light gas mass fractions of less than 10%.

Kepler-62 f is the first known exoplanet whose size (1.4 $R_\oplus$) and orbital position (in the middle of its star's habitable zone) suggests that it could well be a rocky world with stable liquid water at its surface (*Borucki et al.* 2013). Kepler-186 f is a habitable zone planet just a bit larger than Earth (*Quintana et al.* 2014), so it is more likely to be rocky, but because it orbits a smaller, cooler, more variable star, it might not be as good a candidate for habitability as Kepler-62 f. Kepler-452 b is the smallest exoplanet known to orbit within the habitable zone of a G star (*Jenkins et al.* 2015). No mass measurements have been made for any of these potentially habitable exoplanets.

Five of the six planets in the Kepler-80 system are dynamically-coupled in a resonance chain, with neighboring threesomes locked in three-body resonances, i.e., a linear combination of small integer multiples of the orbital frequencies of the planets add to zero. (The outer planet's orbit is not sufficiently well constrained to require a resonant lock, but is consistent with that configuration, whereas the resonance variables of the four planets with better-constrained orbits are clearly librating.) See *MacDonald et al.* (2021) for a detailed analysis of the Kepler-80 system. Kepler-223's four planets are also in a resonance chain, but as these planets have more eccentric orbits, with $e \sim 0.05$ rather than the $e \sim 0.01$ more common for planets in closely-spaced *Kepler* multi-planet systems, the period ratios of neighboring planets are much closer (in this case typically off by $\sim 1$ part in 1000) to small integer ratios (*Mills et al.* 2016; *Lissauer et al.* 2023).

Kepler-88 b is a Neptune-sized planet with an orbital period of just under 11 days that exhibits very large TTVs (12 hour amplitude) and TDVs (Transit Duration Variations, in this case of 5 minute amplitude). These TTVs and TDVs have been used to deduce the presence of its non-transiting jovian-mass companion, Kepler-88 c, which has an orbital period of just over 22 days (*Nesvorný et al.* 2013). Kepler-88 b's TTVs and TDVs are so large because the period ratio of the planets is 2.03, placing them just outside the 2:1 mean motion resonance, and Kepler-88 c is much more massive than other *Kepler* planets that orbit so close to a first-order mean motion resonance. Several other non-transiting planets have been identified and characterized via TTV analysis of transiting planets discovered by *Kepler* (*Nesvorný* 2019). In a far larger number of cases, TTVs imply that a non-transiting planet is present in the system, but the data do not provide enough information to uniquely specify its orbital period. Non-transiting companions to *Kepler* planets have also been detected via radial velocity variations.

Kepler-90 is a star that is slightly larger and more massive than the Sun which hosts eight transiting planets (*Shallue and Vanderburg* 2018), the largest number of planets observed to transit any star. The three inner planets, whose periods range between 7 and 15 days, all have radii $\sim 1.5$ $R_\oplus$. The next three planets, with periods from 59 to 125 days, have radii roughly twice as large as their inner siblings. The outer two planets are both larger than 7 $R_\oplus$ and have orbital periods of 211 and 332 days. While this system has five transiting planets with periods $\gtrsim 60$ days, no other system has more than two known transiting planets on such long-period orbits.

Kepler-444 A is an ancient, low-metallicity star, a bit smaller than the Sun, which hosts five sub-Earth size transiting planets with orbital periods between 3 and 10 days. The stellar age of 11 Gyr, estimated using asteroseismic analysis of stellar oscillations in the *Kepler* lightcurve, is consistent with its galactic orbit and makes this one of the oldest planetary systems known. Each pair of neighboring planets orbit just wide of first-order MMR (i.e., period ratios slightly larger than $(N+1)/N$, where $N$ is a positive integer) with other planets, but no three-body resonance have been identified within the system (*Mills and Fabrycky* 2017). Two low-mass stellar companions, Kepler-444 BC, form a tight spectroscopic binary, whose orbit relative to Kepler-444 A has a semimajor axis of $\sim 50$ au *Zhang et al.* (2023).

Kepler-1520 b produces an asymmetric transit lightcurve when it obscures a portion of its star every 15.7 hours. This lightcurve has been interpreted as a tail of dust being released from a disintegrating planet (*Rappaport et al.* 2012). This disintegrating planet, which can also be thought of as a giant exocomet, is substantially smaller than the Earth.

Several hundred planets have been found using *K2* data. These include small neighbors orbiting both interior (with $P = 0.79$ days) and exterior (with $P = 9.03$ days) to the previously known hot jupiter, WASP-47 b, which has an orbital period $P = 4.16$ days (*Becker et al.* 2015). Thus, this system is an exception to the "rule" that hot jupiters tend to lack close planetary neighbors (*Steffen et al.* 2012a).

The *K2* Mission also detected several small asteroids transiting the white dwarf WD 1145+017 with periods close to 4.5 hours (*Rappaport et al.* 2016). Heavy elements should rapidly sink below the viewable photosphere of white dwarfs, but these elements have been observed in spectra of WD 1145+017 and of more than 25% of the other white dwarfs studied. These polluted white dwarfs must have recently accreted refractory material such as tidally disrupted asteroidal fragments, presumably remnants of the progenitor star's planetary systems.

Despite the difficulty of detecting transiting planets orbiting young stars, which are photometrically noisy, *Kepler* found more than one dozen (young) planets residing in galactic open clusters, with all but two of these discoveries coming from the *K2* phase of the mission (*Rizzuto et al.* 2017). Most of these planets have sizes between those of Earth and Neptune, and all orbit their host stars with periods $P < 26$ days. See *Glaser et al.* (2020) for a listing and discussion of *Kepler*'s planets in clusters.

## 5. PLANET OCCURRENCE RATES

The occurrence frequency and characteristics of small exoplanets orbiting in and near the HZ was completely unknown when *Kepler* was first proposed (1992). It was also recognized that these statistics would be needed for the de-sign of future missions that would characterize planetary atmospheres and search for life. Consequently, the mission was designed to monitor a very large number of stars to obtain useful statistics for characterizing the small planet population in the event that the occurrence rate was low.

Quantifying planet occurrence requires a uniform catalog of planet candidate detections together with quantitative metrics of the survey completeness (the fraction of transiting planets that were identified and vetted as planetary candidates) and reliability (the fraction of planet candidates that are actually planets orbiting the *Kepler* target star). Cumulative (non-uniform) catalogs of planet candidates found using various techniques were released periodically as data analysis from the 4-year monitoring campaign progressed (*Borucki et al.* 2011a,b; *Batalha et al.* 2013; *Burke et al.* 2014; *Rowe et al.* 2015; *Mullally et al.* 2015), each benefiting from a longer dataspan as well as more sophisticated pipeline processing and vetting metrics to weed out astrophysical and instrumental false positives. *Coughlin et al.* (2016) presented the first uniform catalog generated using fully-automated processes from pixels to planet candidate; this was also the first catalog that used all seventeen quarters (4 years) of data acquired on the prime mission field of view. Additional improvements to the pipeline and its characterization were carried out during the extended mission (*Thompson et al.* 2016). In particular, modified data (scrambled light curves, inverted light curves, and light curves with injected transits) were pushed through the same end-to-end pipeline as the actual data in order to tune vetting metrics and quantify the survey completeness and reliability (*Christiansen* 2017; *Coughlin* 2017; *Burke and Catanzarite* 2017a,b).

The *Kepler* Project's final catalog of planet candidates, referred to as the DR25 (Data Release 25) catalog, contains 4034 planet candidates with $0.25 < P < 632$ days (*Thompson et al.* 2018). It is a uniformly-generated catalog optimized for demographic studies. Vetting metrics, which can be adjusted for greater completeness at the expense of reduced reliability or vice versa, were tuned to yield 76.7% vetting completeness and 50.5% vetting reliability for low-S/N planet candidates between 200 and 500 days around FGK-dwarf stars. Dispositions of individual KOIs as planet candidates or false positives can change depending on how the metrics are tuned. A supplemental DR25 catalog release includes the results of ancillary analyses for a subset of targets to produce a "best-knowledge" catalog that prioritized accuracy at the expense of uniformity and reproducibility. (It is not recommended that the supplemental catalog be used for demographic studies.) Parallaxes measured by *Gaia* enabled the calculation of precise star and planet properties from a uniform data set (*Berger et al.* 2018, 2020a), the final ingredients required to produce a fully homogeneous catalog of *Kepler* planetary candidates for statistical inference.

*Lissauer et al.* (2023) provide a unified catalog that incorporates candidates from DR25, DR25supp, previous mission catalogs and those from various other searches as well as *Gaia* constraints on stellar parameters. They perform a reanalysis of *Kepler* lightcurves that improves estimates of various transit parameters, including substantially improved estimates of mean orbital periods during the *Kepler* epoch for PCs exhibiting TTVs. They also revet KOIs with conflicting dispositions among DR24, DR25 and DR25supp. Although they prioritize accuracy over uniformity, they also derive a second set of planetary properties (although not dispositions) that are more accurate than and as uniform as those in DR25, and thus this list of planetary properties is well-suited for use in planet occurrence rate studies.

### 5.1. EARLY RESULTS

Three significant results came out of *Kepler* occurrence studies before the final DR25 catalog was released. Each involved computing an occurrence rate integrated over a specific period-radius domain: 1) hot jupiters, 2) sub-neptunes, and 3) integrated inner planets occurrence.

The hot jupiter (HJ) population was the first to emerge from precision radial velocity (PRV) surveys (*Mayor and Queloz* 1995; *Butler et al.* 1997) owing to the large orbital velocity of the host star. It would take another decade of discovery, however, for sufficient numbers of HJ discoveries (about two dozen) to enable robust estimates of their occurrence rate. Despite the apparent HJ pile-up in the PRV discovery catalogs, such planets are rare, with an average number per FGK field star near 1% (*Marcy et al.* 2005; *Mayor et al.* 2011; *Wright et al.* 2012). Similar analyses using *Kepler* data yield even more extreme results, with HJ occurrences about a factor of two smaller than those from PRV studies (*Howard et al.* 2012; *Santerne et al.* 2012; *Fressin et al.* 2013; *Santerne et al.* 2016; *Petigura et al.* 2018; *Fernandes et al.* 2019; *Kunimoto and Matthews* 2020). The strong dependence of HJ occurrence on host star metallicity (*Santos et al.* 2004; *Fischer and Valenti* 2005) could yield such a discrepancy if *Kepler*'s parent population is metal poor compared to the PRV stars (*Howard et al.* 2012). Indeed, a pile-up of HJ planets can be recovered by considering a subsample of super-metallicity stars from the *Kepler* sample (*Dawson and Murray-Clay* 2013). Moreover, a spectroscopic survey of *Kepler* targets confirms a bias toward lower metallicity (*Guo et al.* 2017). However, the mean metallicity of the *Kepler* sample in that survey is $-0.045 \pm 0.009$ compared to $-0.005 \pm 0.006$ for the PRV sample, and this difference is not large enough to explain the factor of two difference in the HJ occurrence rates. *Guo et al.* (2017) conclude that other factors such as binary contamination (*Moe et al.* 2019) or additional stellar biases must be at play.

Before *Kepler* launched, more than 85% of the exoplanets known were larger than Neptune. In sharp contrast, more than 85% of *Kepler* planet candidates were smaller than Neptune by the third catalog release (*Batalha et al.* 2013). Even in the early planet candidate catalogs, there was a clear mode in the period-radius distribution situated near 2 $R_\oplus$. The first occurrence rate estimates hinted at the high occurrence of planets larger than Earth and smaller than Neptune. Based on the first four months of data, *Borucki et al.* (2011b) reported tantalizing evidence that planets intermediate in size to the terrestrials and giants in the Solar System could outnumber planets with $R_p \sim 1\,R_\oplus$ at orbital period $P < 100$ days. Some of the short-period planets with $1\,R_\oplus < R_p \lesssim 1.6\,R_\oplus$ may be stripped cores of planets that in their youth possessed large (by volume) hydrogen envelopes accreted from the protoplanetary nebula (see review of *Bean et al.* (2021) and references therein).

### 5.2. UPDATED ESTIMATES

As *Kepler* carried out its four-year primary observing campaign, the growing baseline of the photometric time-series enabled the detection of smaller and smaller planets at longer and longer orbital periods. Figure 2 shows occurrence rates of planets around sunlike stars as a function of location in radius-period $(R_p, P)$ space. The binning used for Fig. 2 is (almost) uniform in $\log P$, but non-uniform in $\log R_p$ (as well as $R_p$), which needs to be accounted for when comparing the numbers given within the boxes. However, the colors represent values normalized in differential number in $(\log R_p, \log P)$ space.

Integrating occurrence estimates over the growing domain of high survey completeness in the period-radius plane (e.g., $R_p > 1\,R_\oplus$ and $P < 200$ days), it becomes apparent that the average number of planets per star is at least of order unity for the GK sample (*Youdin* 2011; *Fressin et al.* 2013; *Burke et al.* 2015; *Kunimoto and Matthews* 2020; *Hsu et al.* 2019) and at least twice as large for the early to mid-M dwarf population (*Dressing and Charbonneau* 2015; *Gaidos et al.* 2016; *Hardegree-Ullman et al.* 2019; *Hsu et al.* 2020). This does not, however, imply that every late-type star has a planetary system. More than 20% of the stars with viable candidates in the final (DR25) catalog host multi-planet systems (*Thompson et al.* 2018). Due to this clustering, the fraction of stars with planets, $F_p$, is lower than the average number of planets per star. $F_p$ has been estimated by considering only one planet in each system (*Petigura et al.* 2013a), by forward modeling planet populations via transiting planet detection models (*Lissauer et al.* 2011b; *Mulders et al.* 2018; *He et al.* 2021), and by modeling multiplicity functions based on transit and transit timing distributions (*Zhu et al.* 2018). These methods yield estimates ranging from 30% to more than 60% of GK stars being planet hosts, depending on the method employed and the period-radius domain considered.

### 5.3. RADIUS AND PERIOD DISTRIBUTIONS

Marginalization across the *Kepler* domain to produce radius and period distributions has been fruitful, especially considering small samples in some regions. Figure 3 shows how the mean number of planets per star varies with planetary radius. One of the first-recognized and arguably the most marked feature of the radius distribution is the abrupt drop-off in occurrence between 2.5 and 4 $R_\oplus$ (*Youdin* 2011; *Howard et al.* 2012; *Fressin et al.* 2013; *Dressing and Charbonneau* 2013; *Petigura et al.* 2013a; *Batalha* 2014; *Mulders et al.* 2018; *Kunimoto and Matthews* 2020). *Fulton and Petigura* (2018) and *Hsu et al.* (2019) find that the intrinsic occurrence rate of $2.7 - 3.0\,R_\oplus$ planets is four to ten times that of planets that are only 20% larger ($3.3 - 3.7\,R_\oplus$). This radius cliff is fairly uniform across *Kepler* orbital periods for $P \gtrsim 3$ days. Although planetary abundance per logarithmic interval in $R_p$ continues to increase to smaller sizes (apart from the radius valley discussed below), the rate of increase isn't nearly as large as near the radius cliff (*Hsu et al.* 2019).

*Kite et al.* (2019) proposed that the radius cliff is so steep because for $R_p \gtrsim 3\,R_\oplus$ the pressure at the core/envelope boundary is high enough for envelope to readily dissolve into magma. This sequestration implies that mature planets (i.e., planets that have had time to cool) larger than the radius cliff must retain a substantially larger amount of H/He than do the smaller, more typical, sub-neptunes.

Another striking feature in the marginalized radius distribution (Fig. 3) is the "radius valley", a paucity of planets with $1.5\,R_\oplus \lesssim R_p \lesssim 2\,R_\oplus$ and $P < 100$ days (*Fulton et al.* 2017; *Van Eylen et al.* 2018; see also Fig. 1). This bimodal distribution of radii of short period exoplanets was predicted by theoretical models of envelope mass loss of highly irradiated planets (*Lopez and Fortney* 2013; *Owen and Wu* 2013). Establishing its statistical significance required refinements to *Kepler* host star properties (*Petigura et al.* 2017) as previous uncertainties in stellar radii were obscuring this feature. Mapping planet mass and radius measurements to composition-dependent models suggests that most of the planets above the valley have $1 - 3\%$ mass-fraction hydrogen envelopes; these planets are often referred to as sub-neptunes. In contrast, a significant fraction of planets below the radius valley are consistent with an Earth-like composition (*Lopez and Fortney* 2014; *Rogers* 2015), with rocky planets of $R_p \gtrsim 1.2\,R_\oplus$ often being referred to as super-Earths. The center of the valley shifts to larger planet radius at higher incident flux from the host star (*Van Eylen et al.* 2018; *Martinez et al.* 2019). There is also a discernible shift in the position of the valley towards larger radius around more massive host stars (*Fulton and Petigura* 2018; *Cloutier and Menou* 2020; *Berger et al.* 2020b).

Understanding the physical processes that generate the radius valley has been the focus of intense research (see, for example, *Lopez and Rice* 2018; *Mulders et al.* 2019; *Gupta and Schlichting* 2020; *Lee and Connors* 2021; *Owen and Wu* 2017; *Owen and Campos Estrada* 2020; *Rogers and Owen* 2021; *Rogers et al.* 2021). Three competing models explaining the radius valley have emerged: photoevaporation (*Owen and Wu* 2013, 2017; *Rogers et al.* 2021) and core-powered mass-loss (*Ginzburg et al.* 2016; *Gupta and Schlichting* 2019) both assume H-dominated atmospheres, whereas *Aguichine et al.* (2021) (cf., *Dorn and Lichtenberg* 2021) model the sub-neptunes as having atmospheres dominated by $H_2O$. Despite significantly different mass-loss mechanisms and timescales, both H-dominated atmosphere models can broadly recreate the main characteristics of the radius valley. They yield slightly different locations and shapes of the valley. However, comparisons with *Kepler* occurrence rate distributions have not indicated a statistical preference for one model over the other (*Loyd et al.* 2020). *Berger et al.* (2020b) estimate that a four-fold increase in the sample size of planets will be required to distinguish between physical models using differences in their stellar mass dependencies.

Figure 4 displays variations in the mean number of planets per star with orbital period. The most marked feature in the marginalized period distribution for planets smaller than Neptune is the breakpoint in occurrence rates between $P = 6 - 13$ days that marks the transition between two power laws. Interior to the breakpoint, occurrence (per unit in $\log P$) drops off precipitously and can be explained by formation and migration models. For example, the breakpoint is roughly coincident with the silicate sublimation front (*Flock et al.* 2019) and the co-rotation radius of the disk and stellar magnetosphere resulting in disk truncation (*Lee and Chiang* 2017). Alternatively, the breakpoint can be interpreted as a parking radius in migration scenarios, whereby the innermost planet is trapped at the inner disk edge and halts the migration of other planets (*Cossou et al.* 2014). Beyond the breakpoint, the occurrence distribution flattens out for planets smaller than Neptune (*Dong and Zhu* 2013). However, *Kepler* data do not strongly constrain the abundances of planets with $R_p < 1.5\,\mathrm{R}_\oplus$ and $P > 128$ days (*Hsu et al.* 2019). This region is of particular interest since photoevaporation models predict a paucity of stripped cores beyond $\sim 60$ day orbital periods and a corresponding drop in occurrence for planets in the $1 - 1.6\,\mathrm{R}_\oplus$ size range.

The occurrence rates of large planets also drops precipitously at very short periods, but only for $P < 4$ days. The abundance of planets with sizes between those of Neptune and Saturn gradually rises for longer periods. Jupiter-size planets have a dip in abundance for periods of above 4 days, but start increasing in abundance (again per unit in $\log P$) above $\sim 2$ weeks, as predicted by core accretion models.

### 5.4. DEPENDENCE ON STAR PROPERTIES

The *Kepler* Mission monitored thousands of main sequence stars of each spectral type ranging from F to M, thereby enabling studies of planet occurrence as a function of effective temperature, stellar mass and metallicity. Stellar mass is directly correlated with the masses of protoplanetary disks and inversely correlated with the lifetimes of protoplanetary disks (*Williams and Cieza* 2011). Stellar metallicity is ultimately related to the inventory of solids available in the disk to form planetary cores. In short, star properties and planet occurrence rates are the observable bookends to planet formation and evolution.

Radial velocity surveys revealed a higher occurrence of giant planets around stars of higher mass (*Johnson et al.* 2010) and higher metallicity (*Santos et al.* 2004; *Fischer and Valenti* 2005). The trends suggest that large planet cores are more likely to form in disks with larger reservoirs of solids, as predicted by core-accretion theories (*Lissauer* 1993; *Ida and Lin* 2004; *Mordasini et al.* 2009). Similar studies have now been carried out for the small planet population using *Kepler*'s survey of transiting planets. Running counter to the trend for giant planets, the occurrence of small planets around main sequence stars is inversely correlated with stellar effective temperature and, hence, stellar mass.

Using an early *Kepler* data release, *Howard et al.* (2012) measured the occurrence of planets with radii $2\,\mathrm{R}_\oplus < R_\mathrm{p} < 4\,\mathrm{R}_\oplus$ and periods $< 50$ days orbiting M0 to F2 dwarfs and found these planets are several times as abundant around cool stars (3,600 to 4,100 K) as they are around hot stars (6,600 to 7,100 K). *Mulders et al.* (2015) found the occurrence rate of planets with radii between 1 and 4 $\mathrm{R}_\oplus$ around M dwarfs to be twice as high as for G stars, and three times as high as for F stars, with even larger differences found if the sample is restricted to planets orbiting within 0.1 au of their stellar host (Fig. 5). Most of the M stars in the aforementioned studies were near the upper mass limit for M dwarfs, but *Hardegree-Ullman et al.* (2019) find evidence suggesting that the same trend continues to lower mass (mid-M dwarf) stars.

The larger occurrence rates of small planets around low-mass stars run counter to the trends for giant planets but could be related to the longer disk lifetimes of lower mass stars or the destruction, accretion or removal of small planets by larger ones. It is important to note that *Kepler* occurrence rates are limited to inner orbits (periods less than about one year). Planet formation is a dynamic process that necessitates a systems approach that couples the inner and outer disk. For example, the formation of giant planets outside the snow line can open gaps in the disk and block the inward radial flow of pebbles, preventing the growth of shorter period super-Earths (*Lambrechts et al.* 2019). This can result in an anti-correlation between the occurrence of small inner planets and giant outer planets (*Mulders et al.* 2021).

The relationship between occurrence rates and host star metallicity is not as clear for the small planet population as it is for the large planet population. Early studies compared the metallicity distributions of stars with and without transiting planets. Spectroscopic observations of more than 400 *Kepler* targets confirmed the metallicity trend for giant planets but found a null detection and $3\sigma$ detection of a planet-metallicity correlation for planets with radii in the range $1\,\mathrm{R}_\oplus < R_\mathrm{p} < 1.7\,\mathrm{R}_\oplus$ and $1.7\,\mathrm{R}_\oplus < R_\mathrm{p} < 4\,\mathrm{R}_\oplus$, respectively (*Buchhave and Latham* 2015). The authors concluded that small planets ($R_\mathrm{p} < 4\,\mathrm{R}_\oplus$) form around host stars with a wide range of metallicities. *Wang and Fischer* (2015) however, reanalyzed the data after removing known systematics from the photometrically-derived star properties and concluded that the occurrence rate of planets with $R_\mathrm{p} < 4\,\mathrm{R}_\oplus$ in their metal-rich group is about twice as high as that in their metal-poor group.

Later studies compared planet occurrence for high and low metallicity samples. *Mulders et al.* (2016) concluded that short-period (less than 10 days) exoplanets are preferentially found around metal-rich stars. No correlation is found for small planets at longer orbital periods. *Petigura et al.* (2018) perform a similar study for planets with periods between 10 and 100 days. The authors find no correlation between occurrence and metallicity for terrestrial-size planets ($R_p < 1.7 \, R_\oplus$). However, the occurrence of sub-neptunes ($1.7 \, R_\oplus < R_p < 4 \, R_\oplus$) orbiting metal-rich stars is double the occurrence of these larger planets orbiting metal-poor stars. Leveraging the large LAMOST spectroscopic survey of *Kepler* targets, *Zhu* (2019) studies occurrence of $P < 400$ day and $R_p > 1 \, R_\oplus$ planets expressed as both the average number of planets per star and the fraction of stars with planets. They confirm that the formation efficiency of small planets depends on stellar metallicity, but they also conclude that the dependence is much weaker than for giant planets.

### 5.5. ESTIMATES OF $\eta_\oplus$

Quantifying the mean number of Earth analog planets per star, denoted $\eta_\oplus$, is of great interest to astrobiologists. Prior to *Kepler*, estimates of $\eta_\oplus$ were based primarily on theoretical models of planet formation because there were no exoplanet data on planets like Earth. *Kepler* has contributed to estimating $\eta_\oplus$ in two ways: measuring the differential abundance (denoted $\Gamma(R_p, P)$) of planets near the position of Earth in the radius-period plane (Fig. 2) and helping understand the maximum size at which planets are likely to be rocky. *Rogers* (2015) showed that most planets with radii $R_p \geq 1.6 \, R_\oplus$ are not rocky, but that result is based primarily on measurements of the masses and radii of planets with periods shorter than one month, and since the radius valley appears at smaller sizes for longer period planets (Fig. 1), the upper limit for planets considered to be likely Earth analogs should be $\sim 1.5 \, R_\oplus$ or smaller.

Early estimates of $\eta_\oplus$ based upon *Kepler* data (*Youdin* 2011; *Petigura et al.* 2013b; *Foreman-Mackey et al.* 2014; *Burke et al.* 2015) varied widely owing to factors such as the state of the survey (incomplete data and/or poorly characterized survey completeness and/or reliability; of special importance are "rolling band" artifacts (*Bryson et al.* 2020) generating false positives roughly commensurate with the spacecraft orbital period at 372 days), systematic errors in star and planet properties, the size range adopted to describe rocky planets, and the varying definitions of the habitable zone (see summaries by *Burke et al.* 2015; *Kunimoto and Matthews* 2020). *Raymond et al.* (2007) considered 0.3 $M_\oplus$ as the lower-mass limit for planetary habitability since smaller planets are less likely to retain the substantial atmospheres and ongoing tectonic activity that could be required to support life. This mass corresponds to 0.69 $R_\oplus$ using the mass-radius relation of *Sotin et al.* (2007). Compositional analysis of planets in the mass-radius plane (Figure 6) suggests a transition between Earth-like and volatile-rich compositions near 1.5 $R_\oplus$ (*Rogers* 2015), which lies near the lower edge of the radius valley (*Fulton et al.* 2017).

In 2017, the Study Analysis Group (SAG) 13 of the NASA Exoplanet Exploration Program Analysis Group (ExoPAG) recommended adopting standardized definitions for planet yield estimates in future mission concept studies, namely: $0.5 < R_p < 1.5 \, R_\oplus$ and $237 < P < 860$ days, the latter corresponding to the optimistic habitable zone for sunlike stars defined by *Kopparapu* (2013). This, combined with the availability of the final DR25 survey products and *Gaia* star properties, allowed for better estimates of $\eta_\oplus$.

An assessment of the full habitable zone requires extrapolation from data yielding robust detections at shorter orbital periods and/or larger planet radii. The noise levels attained by *Kepler* on-orbit exceed the levels expected at time of launch for quiet solar-type stars by $\sim 50\%$, due primarily to intrinsic stellar variability on time scales similar to transit durations (*Gilliland et al.* 2011). With the loss of a second reaction wheel in 2013, there was no recourse for recovering the baseline sensitivity via continued observations of the primary field in an extended mission.

Analysis efforts based on DR25 survey products (*Mulders et al.* 2018; *Hsu et al.* 2019; *Zink and Hansen* 2019) yield comparable $\eta_\oplus$ values ranging from 0.2 to 0.67 for GK stars (Earth-analogs would be even more difficult to detect around F stars, since the stellar radii are larger and the orbital periods of planets in the HZ are longer), though careful treatment of the catalog reliability due to rolling band artifacts is required (*Bryson et al.* 2020). Incorporating all reliability metrics yields $0.37^{+0.48}_{-0.21} - 0.60^{+0.90}_{-0.36}$, depending on the type of extrapolation employed (*Bryson et al.* 2021).

The estimated occurrence rates for Earth-analog planets orbiting GK dwarfs are in rough agreement with similar $\eta_\oplus$ calculations for M dwarfs: $0.33^{+0.10}_{-0.12}$ for $0.75 < R_p < 1.5 \, R_\oplus$ (*Hsu et al.* 2020) and $0.16^{+0.17}_{-0.07}$ for a more limited radius range $1.0 < R_p < 1.5 \, R_\oplus$ (*Dressing and Charbonneau* 2015). Planets in the habitable zone of M dwarfs have much shorter orbital periods than do the planet candidates impacted by rolling band artifacts, and 1 $R_\oplus$ planets produce much deeper transits when orbiting an M dwarf rather than a sunlike star. Unlike for the GK population, direct calculation of $\eta_\oplus$ (at least down to 1 $R_\oplus$) for M dwarfs is possible without the need for extrapolation. Regarding the GK population, however, *Lopez and Rice* (2018) and *Pascucci et al.* (2019) caution that the necessary extrapolations from a period-radius domain that could include sizable numbers of stripped cores can lead to overestimated occurrence rates of Earth analogs.

### 6. PLANETARY BULK PROPERTIES AND COMPOSITIONS

Measurements of both the size and mass of a planet provide constraints on its bulk composition. Masses are not directly measurable from *Kepler* data, except when they can be derived from TTVs and a few rare cases of massive plan-ets on very short-period orbits for which $M_p$ can be measured from the *Kepler* lightcurve (see Sec. 7). Nonetheless, the majority of the sub-saturn exoplanets whose masses and radii are both known are *Kepler* discoveries, although in recent years most of the newly-added planets have been found by *TESS*. The Doppler technique that uses ground-based spectroscopic observations to detect variations of the host star's radial velocity in response to the gravitational tugs of the planet has provided many measurements of the masses of *Kepler* planets. Especially significant among these are mass measurements of numerous short-period planets with $R_p < 2.5$ R$_\oplus$ that orbit relatively bright (for *Kepler*) stars (*Marcy et al.* 2014). Analysis of TTVs evident in some *Kepler* lightcurves has provided measurements of masses of sub-neptune mass *Kepler* planets with orbital periods of up to several months (e.g., *Hadden and Lithwick* 2017; *Jontof-Hutter et al.* 2021).

The density of a planet, easily calculable from its mass and size, provides a good single-parameter approximation to compositional constraints for planets almost exclusively composed of liquids and solids, although corrections must be made for the self-compression that causes density to increase slowly with radius when material of similar bulk composition is added to a planet (e.g., *Valencia et al.* 2007). Kepler-36 b is a particularly interesting example, because *Kepler* photometry allows the ratio $R_p/R_\star$ to be computed very precisely, and strong TTVs induced in its proximate neighbor enable low-uncertainty estimates of $M_p/M_\star$. The density of the host star, Kepler-36, is better constrained by asteroseismic oscillations observed by *Kepler* than are $M_\star$ and $R_\star$ individually, resulting in correlated uncertainties in the mass and radius of the star, and by extension that of the planet (*Carter et al.* 2012). Taking into account these correlations allows the core mass fraction of Kepler-36 b (assuming this planet's core and mantle to have the same compositions as the corresponding regions of Earth) to be constrained to be a value very similar to that of our planet, whereas significantly weaker compositional constraints are obtained by accounting for the uncertainties of $M_p$ and $R_p$ independently.

In sharp contrast, density is a far poorer indicator of composition than radius alone for most known planets that are primarily gaseous by volume, with the stellar flux and age being as or more important than mass for a given radius (*Lopez and Fortney* 2014). Note that a tenuous H-He or pure H atmosphere, even if it accounts for as little as 0.1% of the planet's total mass, can occupy most of the planet's volume and make a huge difference to the planetary radius. Saturn-sized and larger planets must be primarily hydrogen by volume. Even Neptune-size planets must have H and/or He filling at least tens of percent of their volumes unless they are extraordinarily dense.

Figure 6 shows the mass-radius-temperature relationship for planets with precise measurements of both mass and radius that were discovered by the *Kepler* spacecraft, including the *K2* phase of the mission. The smallest exoplanets shown are true super-Earths, in the restricted sense of being significantly larger than Earth and having densities that strongly suggest a composition at least grossly similar to that of our home planet. Most have orbital periods of less than two weeks. The prevalence of such rocky compositions implies that most of these hot planets formed interior to the ice condensation line within their protoplanetary disk, although it does not preclude inward migration by up to $\sim 1$ au.

In contrast, most exoplanets with well-constrained masses and radii $R_p > 1.7$ R$_\oplus$ are less massive than Earth-composition planets their size would be, implying that they contain volumetrically-significant amounts of less dense constituents. This suggests a boundary between rocky and "other" planets at $\sim 1.6$ R$_\oplus$ (*Rogers* 2015). Unfortunately, data are largely restricted to planets subjected to high incident stellar flux (and thus having high equilibrium temperature, $T_{\rm eq}$) and are likely to suffer from detectability biases, so stronger conclusions cannot be made at present. Moreover, the situation is clearly more complex than mass being a single-valued function of planetary size. Kepler-11 b is near the upper boundary of the "super-Earth-size" range, and has a mass implying a substantial (by volume) component of lighter material, which could be $H_2O$ and/or H/He. Many other planets, including four of this planet's compatriots Kepler-11 c-f, lie significantly above the pure water curve in the mass-radius plane, and each must have a significant fraction of its volume filled with H/He (*Lissauer et al.* 2013).

Indeed, most planets with $2$ R$_\oplus < R_p < 8$ R$_\oplus$ and $P < 100$ days have quite low density and lie above the 50% water/50% rock curve in Fig. 6, indicating voluminous H/He envelopes (*Rogers et al.* 2011; *Lopez and Fortney* 2014), although near the small end of this range, i.e., $\sim 2 - 3$ R$_\oplus$, many could be water worlds with $H_2O$ and perhaps other astrophysical "ices" filling most of their volumes (*Aguichine et al.* 2021). Heavier elements must comprise the vast majority of the *masses* of most of these large and light worlds, but some planets have extremely low densities, implying extended H/He envelopes that dominate by *volume*. High-altitude photochemical hazes, such as those invoked to explain the large radii of super-puff planets by *Gao and Zhang* (2020) (see Sec. 4), add to the ambiguity of deducing composition of planets with weakly-bound atmospheres from mass, radius and impinging stellar radiation flux information alone. Nonetheless, the large numbers of planets with low-mass H/He envelopes compared to massive giant planets can be easily understood in the framework of core-accretion models (*Pollack et al.* 1996; *Bodenheimer et al.* 2018), and is seen as supporting evidence of such formation pathways.

The presence of the radius valley, the paucity of *Kepler* planets with $R_p \sim 1.7$ R$_\oplus$ compared to somewhat smaller and somewhat larger planets (e.g., *Berger et al.* 2020b), can be explained by the loss of H/He envelopes of rock+gas planets. Alternatively, the dissolution of water into a magma ocean may provide an explanation for this dip in the abundance of planets with $R_p \sim 1.7$ R$_\oplus$, if water-rich (several percent to tens of percent by mass) and H$_2$+He-poor inner planets are common (*Kite and Schaefer* 2021).

Although most transiting hot jupiters were discovered using ground-based photometric surveys, *Kepler* found the majority of transiting giant planets with periods longer than two weeks that were known prior to the recent finds of the *TESS* spacecraft. As predicted by models of cool massive H/He-dominated planets, but in sharp contrast to hot jupiters (many of which are more than twice as voluminous as Jupiter), none of the well-studied members of this population have $R_p > 1.2\,R_{\rm Jupiter}$.

## 7. PHASE CURVES & OCCULTATIONS

Phase curves (light curves folded and summed at the planet's orbital period) and occultations (also referred to as secondary eclipses) provide an abundance of information about the characteristics of some short-period *Kepler* planets beyond that gained from the transits alone. In particular, estimates of the planets' albedo, brightness temperature, the presence of an atmosphere, variations in cloud cover, and winds, as well as the mass of the planet, have been obtained (*Mazeh et al.* 2012; *Demory* 2014; *Esteves et al.* 2015). The timing of the occultation provides constraints on orbital eccentricity. The depth of the occultation gives information about the albedo and daytime temperatures. The longitudinal variations of the planetary emission provide information on the winds and clouds. Masses of close-in super-jupiters can often be derived from both ellipsoidal variations of the host star and Doppler boosting of starlight (brightening when moving towards the observer and becoming fainter when receding, also referred to as Doppler beaming).

One of *Kepler*'s first discoveries was the detection of reflected light and thermal emission from the exoplanet HAT-P-7 b (*Borucki et al.* 2009). The measurements of the phase curve and occultations for HAT-P-7 were followed up by similar analyses of Kepler-4 b, Kepler-5 b, Kepler-6 b, Kepler-7 b, Kepler-8 b, Kepler-10 b, Kepler-12 b, Kepler-13 b, Kepler-41 b, Kepler-76 b, Kepler-91 b, and Kepler-412 b (*Désert et al.* 2011; *Esteves et al.* 2015; *Sheets and Deming* 2017). *Angerhausen et al.* (2015) find that in a sample of 18 confirmed hot jupiters, all with periods $1.4 < P < 7$ days, most have low albedos ($< 0.1$), although a few having albedo values exceeding 0.2.

The mass of Kepler-13 b, a very massive hot jupiter orbiting a star that is too hot ($T_{\rm eff} \approx 7500$ K) for precise RV measurements, was measured directly from its host star's lightcurve, which shows signatures both of Doppler beaming as a result of the star's orbit about the star-planet center-of-mass and the tidal distortion of the stellar photosphere that the planet's gravity induces (*Mazeh et al.* 2012). *Faigler and Mazeh* (2015) and *Esteves et al.* (2013) made similar measurements of several more *Kepler* hot jupiters.

Although derivation of planet masses from phase curves is generally only possible for giant planets orbiting close to their host stars, signal-to-noise ratios from some small planets are sufficient to determine brightness temperatures and albedos. An analysis by *Demory* (2014) of the phase curves for 97 planetary candidates with radii $R_p < 2\,R_\oplus$ provided estimates of planetary brightness temperatures and albedos for 27 candidates. The average albedos, 0.16 to 0.30, are about triple those for hot jupiters (0.06 to 0.11). Examination of 56 candidates with $R_p < 2\,R_\oplus$ by *Sheets and Deming* (2017) found a somewhat lower value of $0.11 \pm 0.06$ after excluding Kepler-10 b with its albedo of 0.6 (*Sheets and Deming* 2014).

In the 2000 proposal to NASA's Discovery program, *Kepler* was predicted to discover $\sim 870$ non-transiting hot jupiters via phase curve variations. Since that time, the albedos of many transiting hot jupiters have been measured and found to be typically much smaller than those of Jupiter and Saturn ($\sim 0.5$), on which the forecast was based. In addition to being less reflective, the estimated occurrence rate of hot jupiters has dropped by a factor of a few (*Howard et al.* 2012). Nonetheless, *Millholland and Laughlin* (2017) identified 60 candidate non-transiting hot jupiters in the *Kepler* data. *Lillo-Box et al.* (2021) confirmed three of these candidates using RV observations, and provided less conclusive supporting evidence for a few additional candidates.

## 8. PLANETARY SYSTEMS

The focus of the spacecraft's design was to search for small planets in their host star's habitable zone, but *Kepler*'s ultra-precise, long duration photometry is also ideal for detecting systems with multiple transiting planets (multis). *Kepler* discovered more than 700 candidate multis (see Figures 1 and 7), almost all of which represent true multiple planet systems (*Lissauer et al.* 2011b; *Fabrycky et al.* 2014). The multi-transiting systems from *Kepler* provide a large and very rich data set of short-period (primarily from two days to several months) planets that can be used to reveal the dynamical architectures of exoplanetary systems and to powerfully constrain theoretical models of the formation and evolution of planetary systems (*Lissauer et al.* 2011b; *Tremaine and Dong* 2012; *Fabrycky et al.* 2014; *Lissauer et al.* 2014a; *Winn and Fabrycky* 2015; *He et al.* 2020, 2021; *Millholland et al.* 2021; *Lissauer et al.* 2023, Lissauer et al. in prep.).

Dynamical modelling has been used to explore the typical spacings of planets, orbital resonances within planetary systems, and relative inclinations of planets. Correlations have been found between properties of planets within the same system as well as between planetary properties and system architectures. Detection biases are often present and need to be accounted for to understand the statistical significance of the results. However, for most questions these biases are not as severe as they are for computing occurrence rates (Sec. 5). Thus, many studies sacrifice sample uniformity in order to increase the size of the sample and thereby reduce stochastic uncertainties (*Lissauer et al.* 2023).

Forward modeling uses theoretical distributions of planetary system characteristics in three dimensions contain-ing one or more free parameters to simulate transit yields. These predictions are then compared with the observed sample to fit the parameters. Such forward models have been very helpful in elucidating various planetary system properties such as typical number of planets (within a specified range in size and period) per system and fraction of stars possessing planets (*Lissauer et al.* 2011b; *Zhu et al.* 2018; *He et al.* 2019, 2020, 2021).

Major findings of the studies of *Kepler* multis include:

1) The large number of candidate multiple transiting planet systems observed by *Kepler* shows that flat (i.e., nearly coplanar) multi-planet systems are common. Planets smaller than $\sim 4$ R$_\oplus$ are more likely to be in such flat systems than are larger planets (*Latham et al.* 2011; *Lissauer et al.* 2023).

2) The inclination dispersion of candidate systems with two or more transiting planets has a median of $\lesssim 2°$ (e.g., *Fabrycky et al.* 2014; *He et al.* 2021), suggesting that (inner) planetary systems with relatively low mutual orbital inclinations similar to those of planets within the Solar System are common.

3) Orbits of most planets in multi-transiting systems are nearly circular, with typical eccentricities of only a few percent. This ellipticity is similar to our Solar System, but small compared to those of most giant exoplanets with orbital periods of more than ten days. Larger eccentricities are common among *Kepler* planets that do not have observed transiting companions (*Xie et al.* 2016; *He et al.* 2020), and planets in high-multiplicity systems typically are on more circular orbits than those in multis containing just two transiting planets (*Lissauer et al.* 2023).

4) Most planets found by *Kepler* are neither in nor especially close to first-order mean motion resonances with other planets, but there is a statistically-significant excess of planet pairs spaced slightly wide of resonance, and a deficit of planet pairs whose orbital period ratio is slightly smaller than the resonant ratio (Figure 8). Resonant chains involving three or more planets, while unusual, are far more common than can be explained by chance distributions of periods, implying that substantial convergent migration occurred, at least in a sub-population of inner planetary systems (*Lissauer et al.* 2011b). Excluding planets in resonance chains, the overabundance of planet pairs with period ratio slightly larger than that of first-order MMR approximately cancels the under abundance of pairs with period ratio a little less than first-order MMR (Lissauer et al., in prep.).

5) Attempts to generate a simulated ensemble of planetary systems to match the observed *Kepler* planets have difficulty doing so using a single homogeneous population. In particular, there is evidence for a population or subpopulation of systems that either contains only one detectable planet per star or multiple planets with high mutual orbital inclination. Another population, whose characteristics are described in item 7, can account for the vast majority of multi-transiting systems, as well as a significant minority of the single planets observed (*Lissauer et al.* 2011b; *He et al.* 2021). A third, rarer, group of planar densely-packed multi-planet systems, might also be present. Note that these populations of planetary systems need not be cleanly separated, i.e., there may be significant numbers of intermediate systems (*Millholland et al.* 2021). The added difficulty in detecting multiple planets around the same stellar host suggests that the population of multi-transiting systems may be larger relative to those of singly-transiting systems than estimated by studies that ignored this detection bias (*Zink and Hansen* 2019).

6) Roughly half of solar-type stars have one or more planets in the size range 0.5 R$_\oplus < R_p < 10$ R$_\oplus$ with periods $3 < P < 300$ days. The fraction of stars with such planets increases with decreasing stellar size from spectral type F2V to mid-K dwarfs (*He et al.* 2021).

7) The abundance and distribution of singly and multiply transiting systems implies that $\sim 30\%$ of sunlike stars have one or more planets larger than Earth ($R_p > $ R$_\oplus$) and orbital period $P < 400$ days. The mean number of such planets per star for those stars with at least one such planet is $\sim 3$ (*Zhu et al.* 2018).

8) Many *Kepler* targets with one or more identified planet candidate(s) must be multi-planet systems where additional planets are present that are not transiting and/or too small to have been detected. The presence of some of these unseen companions is betrayed by patterns of TTVs in the transiting planets that are not attributable to observed companions. Nonetheless, TTVs have been detected for a significantly larger fraction of planets in multis than in single-transiting planet systems (*Holczer et al.* 2016; *Lissauer et al.* 2023).

9) Systems with more than one planet having orbital period $P < 2.4$ days are quite rare. Kepler-42, a very small star ($R_\star \lesssim 0.2$ R$_\odot$, *Muirhead et al.* 2012), is the single known exception, and hosts three such planets. Note that neither *Kepler* nor most other exoplanet surveys target a large number of very low mass stars (because most such stars are extremely faint, especially at visible wavelengths), so multiple very short-period planets might not be unusual around very small stars.

10) Large angles between the direction of the angular momentum of planetary orbits and that of stellar spin (known as stellar obliquities), which have been commonly found for hot jupiters orbiting stars with $T_{\rm eff} > 6200$ K, are uncommon for stars hosting small transiting planets, especially hosts of multi-transiting systems (*Winn et al.* 2017). However, there are exceptions such as Kepler-56, which hosts two co-planar transiting planets with orbital periods $P > 10$ days, yet has a stellar obliquity well above $30°$.

11) Sizes of planets within a given system are correlated. In other words, planets orbiting the same host star tend to be more similar in size to each other than randomly-drawn groups of planets (*Weiss et al.* 2018).

## 9. LOOKING FORWARD

NASA's *Kepler* Mission vastly expanded humanity's knowledge of exoplanets and exoplanetary systems within $\sim 1$ AU of their host stars. *Kepler* data show that the average number of planets per star exceeds unity and that flat inner multi-planet systems are common. These data have also substantially enhanced our understanding of stellar properties through asteroseismic analysis of short-period oscillations (e.g., *Huber et al.* 2017) and long-term measurements of stellar variability.

Analysis of the *Kepler* data used in the search for planets, especially the smallest ones, is quite complex and time-consuming. Thus, data analysis and interpretation will continue for decades to come. New planetary candidates will be identified, many more candidates will be verified as planets, and more of the chaff (false positives) will be removed from the candidate pool. And many *Kepler* planets will be characterized better through future observations and/or analysis of TTVs and stellar properties.

*Kepler*, while exceptionally productive, did not complete the task of providing a direct measurement of the abundance of Earth-size planets in the habitable zones of sunlike stars. Continued analysis of the data collected by *Kepler*, including better understandings of the losses in detection efficiency due to both stellar and planetary multiplicity, will help refine estimates of $\eta_\oplus$, as will studies that investigate the false positive rate of low S/N planet candidates most similar to Earth. However, without new transit survey data, the uncertainty will be dominated by the need for extrapolation to obtain a statistically-significant sample size. (At present, no exoplanets with size, host star size and orbital period all within 30% of Earth are known, and even if some *Kepler* planet candidates that occupy this region of parameter space are eventually verified, the number of such planets will be very small.)

The most rapid and cost-effective way to complete this key science goal would be to launch a *Kepler*-like spacecraft (same telescope design, but improved electronics and reaction wheels) and use it to provide additional observations of the same FOV *Kepler* pointed at during its prime mission. The joint analysis of the lightcurves from the two missions would substantially increase the detectability of the Earth-size planets in the HZ of sunlike stars and thereby provide the data necessary to determine the occurrence rate of such planets. NASA's *TESS* spacecraft has recorded lightcurves of *Kepler* target stars, yielding new transit time data for TTV studies of various large planets found by *Kepler* as well some smaller planets around brighter stars and also a wealth of data on decadal time scale stellar variability. However, with only $\sim 1\%$ of *Kepler*'s light-gathering power, *TESS* data are not helpful for finding Earth-analogs. Data from ESA's *PLATO* spacecraft (*Rauer et al.* 2014), scheduled for launch in 2026, may be of some use for computing planet occurrence rates if it follows the current observing plan to place $\sim 30\%$ of the *Kepler* FOV in its all-camera overlap region for two years (*Nascimbeni et al.* 2022), this will enable it to collect half as many photons from a transit as *Kepler*, providing potentially useful data for a joint lightcurve search. Increasing the overlap and/or the amount of time it observes this FOV will improve the chances of substantially augmenting the *Kepler* data useful for estimating $\eta_\oplus$, although such data would still not be as useful as data from a second *Kepler* would be.

Analogs to our Solar System are unlikely to have been detected by *Kepler*. All of the Sun's planets are smaller than planet candidates that have been identified at similar orbital periods using *Kepler* data. As data analysis continues, Venus analogs may well be found, and perhaps Earth analogs, albeit only around the brightest and least variable stars. Giant planet transits would be seen around most *Kepler* targets if the geometry is right. But fewer than 1 in 2000 randomly-positioned distant observers of the Solar System would see even a single giant planet transit during the interval in which *Kepler* took data; therefore, even if giant planet configurations like our Solar System are common, only $\sim 80$ such transits should be present in the *Kepler* data. A few dedicated searches have found several long-period planet candidates (e.g., *Kawahara and Masuda* 2019), but as no systematic studies of the detection efficiencies of such efforts have been made, results cannot be used to make reliable estimates of the occurrence rates of long-period giant planets.

The National Academy of Sciences' 2020 Astronomy Decadal Survey *Pathways to Discovery in Astronomy and Astrophysics for the 2020s* recommends a future flagship NASA mission to image an Earth-like planet. The estimated numbers of planets that such a mission with a specified set of parameters would yield hinged on planetary occurrence rates measured by *Kepler*.

By almost any measure, *Kepler* was one of NASA's most successful scientific discovery missions, the more so if compared with other missions of comparable cost. Ten years after its launch, the majority of planets known to humanity were *Kepler* discoveries. More importantly, *Kepler* vastly expanded the parameter space of known exoplanet characteristics, and showed that the number of planets in our galaxy exceeds the number of stars.

**Acknowledgments** We thank Doug Caldwell for answering some questions that we had about the downloaded *Kepler* data, David Stevenson for providing theoretical insight on the planetary mass-radius relationship, and Jeff Cuzzi for numerous constructive comments on a draft version of this manuscript.

The *Kepler* Mission was the tenth Discovery Mission. The work was supported by the NASA Headquarters Science Mission Directorate. The NASA Ames Research Center and Jet Propulsion Laboratory managed the Mission. The Science Office was located at NASA Ames. The Ball Aerospace and Technology Corporation built the instrument and spacecraft.

**REFERENCES**

Agol E. and Carter J. A., 2018 *NewAR*, *83*, 18.
Aguichine A. et al., 2021 *ApJ*, *914*, 2, 84.
Angerhausen D. et al., 2015 *PASP*, *127*, 957, 1113.
Barclay T. et al., 2013 *Nature*, *494*, 7438, 452.
Batalha N. M., 2014 *Proc. National Academy Science*, *111*, 35, 12647.
Batalha N. M. et al., 2010a *ApJL*, *713*, 2, L103.
Batalha N. M. et al., 2010b *ApJL*, *713*, 2, L109.
Batalha N. M. et al., 2011 *ApJ*, *729*, 1, 27.
Batalha N. M. et al., 2013 *ApJS*, *204*, 2, 24.
Bean J. L. et al., 2021 *Journal of Geophysical Research (Planets)*, *126*, 1, e06639.
Becker J. C. et al., 2015 *ApJL*, *812*, 2, L18.
Berger T. A. et al., 2018 *ApJ*, *866*, 2, 99.
Berger T. A. et al., 2020a *AJ*, *159*, 6, 280.
Berger T. A. et al., 2020b *AJ*, *160*, 3, 108.
Bodenheimer P. et al., 2018 *ApJ*, *868*, 2, 138.
Borucki W. et al., 2018 *NewAR*, *83*, 28.
Borucki W. J., 2016 *Reports on Progress in Physics*, *79*, 3, 036901.
Borucki W. J. et al., 2003 *Future EUV/UV and Visible Space Astrophysics Missions and Instrumentation.*, vol. 4854 of *Society of Photo-Optical Instrumentation Engineers (SPIE) Conference Series* (J. C. Blades and O. H. W. Siegmund), pp. 129–140.
Borucki W. J. et al., 2009 *Science*, *325*, 5941, 709.
Borucki W. J. et al., 2010 *Science*, *327*, 5968, 977.
Borucki W. J. et al., 2011a *ApJ*, *728*, 117.
Borucki W. J. et al., 2011b *ApJ*, *736*, 1, 19.
Borucki W. J. et al., 2013 *Science*, *340*, 6132, 587.
Brown T. M., 2003 *ApJL*, *593*, L125.
Brown T. M. et al., 2011 *AJ*, *142*, 4, 112.
Bryson S. et al., 2020 *AJ*, *160*, 5, 200.
Bryson S. et al., 2021 *AJ*, *161*, 1, 36.
Bryson S. T. et al., 2013 *PASP*, *125*, 930, 889.
Buchhave L. A. and Latham D. W., 2015 *ApJ*, *808*, 2, 187.
Burke C. J. and Catanzarite J., 2017a Planet Detection Metrics: Per-Target Detection Contours for Data Release 25, Kepler Science Document KSCI-19111-002.
Burke C. J. and Catanzarite J., 2017b Planet Detection Metrics: Per-Target Flux-Level Transit Injection Tests of TPS for Data Release 25, Kepler Science Document KSCI-19109-002.
Burke C. J. et al., 2014 *ApJS*, *210*, 2, 19.
Burke C. J. et al., 2015 *ApJ*, *809*, 1, 8.
Butler R. P. et al., 1997 *ApJL*, *474*, 2, L115.
Caldwell D. A. et al., 2010 *ApJL*, *713*, 2, L92.
Carter J. A. et al., 2012 *Science*, *337*, 6094, 556.
Chaplin W. J. et al., 2014 *ApJS*, *210*, 1, 1.
Charbonneau D. et al., 2000 *ApJL*, *529*, 1, L45.
Christiansen J. L., 2017 Planet Detection Metrics: Pixel-Level Transit Injection Tests of Pipeline Detection Efficiency for Data Release 25, Kepler Science Document KSCI-19110-001.
Cloutier R. and Menou K., 2020 *AJ*, *159*, 5, 211.
Cochran W. D. et al., 2011 *ApJS*, *197*, 7.
Cossou C. et al., 2014 *A&A*, *569*, A56.
Coughlin J. L., 2017 Planet Detection Metrics: Robovetter Completeness and Effectiveness for Data Release 25, Kepler Science Document KSCI-19114-002.
Coughlin J. L. et al., 2016 *ApJS*, *224*, 1, 12.
Dawson R. I. and Murray-Clay R. A., 2013 *ApJL*, *767*, 2, L24.
Demory B.-O., 2014 *ApJL*, *789*, 1, L20.
Désert J.-M. et al., 2011 *ApJ*, *197*, 1, 14.
Désert J.-M. et al., 2015 *ApJ*, *804*, 1, 59.
Dong S. and Zhu Z., 2013 *ApJ*, *778*, 1, 53.
Dorn C. and Lichtenberg T., 2021 *ApJL*, *922*, 1, L4.
Doyle L. R., 2019 *NewAR*, *84*, 101515.
Dressing C. D. and Charbonneau D., 2013 *ApJ*, *767*, 1, 95.
Dressing C. D. and Charbonneau D., 2015 *ApJ*, *807*, 1, 45.
Esteves L. J. et al., 2013 *ApJ*, *772*, 1, 51.
Esteves L. J. et al., 2015 *ApJ*, *804*, 2, 150.
Fabrycky D. C. et al., 2012 *ApJ*, *750*, 114.
Fabrycky D. C. et al., 2014 *ApJ*, *790*, 2, 146.
Faigler S. and Mazeh T., 2015 *ApJ*, *800*, 1, 73.
Fernandes R. B. et al., 2019 *ApJ*, *874*, 1, 81.
Fischer D. A. and Valenti J., 2005 *ApJ*, *622*, 2, 1102.
Flock M. et al., 2019 *A&A*, *630*, A147.
Ford E. B. et al., 2012 *ApJ*, *750*, 113.
Foreman-Mackey D. et al., 2014 *ApJ*, *795*, 1, 64.
Fressin F. et al., 2012 *Nature*, *482*, 7384, 195.
Fressin F. et al., 2013 *ApJ*, *766*, 2, 81.
Freudenthal J. et al., 2018 *A&A*, *618*, A41.
Fulton B. J. and Petigura E. A., 2018 *AJ*, *156*, 6, 264.
Fulton B. J. et al., 2017 *AJ*, *154*, 3, 109.
Gaia Collaboration et al., 2018 *A&A*, *616*, A1.
Gaidos E. et al., 2016 *MNRAS*, *457*, 3, 2877.
Gao P. and Zhang X., 2020 *ApJ*, *890*, 2, 93.
Gilliland R. L. et al., 2011 *ApJS*, *197*, 1, 6.
Ginzburg S. et al., 2016 *ApJ*, *825*, 1, 29.
Glaser J. P. et al., 2020 *AJ*, *160*, 3, 126.
Guo X. et al., 2017 *ApJ*, *838*, 1, 25.
Gupta A. and Schlichting H. E., 2019 *MNRAS*, *487*, 1, 24.
Gupta A. and Schlichting H. E., 2020 *MNRAS*, *493*, 1, 792.
Hadden S. and Lithwick Y., 2017 *AJ*, *154*, 1, 5.
Hardegree-Ullman K. K. et al., 2019 *AJ*, *158*, 2, 75.
He M. Y. et al., 2019 *MNRAS*, *490*, 4, 4575.
He M. Y. et al., 2020 *AJ*, *160*, 6, 276.
He M. Y. et al., 2021 *AJ*, *161*, 1, 16.
Henry G. W. et al., 1999 *IAUC*, *7307*, 1.
Holczer T. et al., 2016 *ApJS*, *225*, 9.
Holman M. J. and Murray N. W., 2005 *Science*, *307*, 5713, 1288.
Holman M. J. et al., 2010 *Science*, *330*, 51.
Howard A. W. et al., 2012 *ApJS*, *201*, 2, 15.
Howell S. B. et al., 2021 *Frontiers in Astronomy and Space Sciences*, *8*, 10.
Hsu D. C. et al., 2019 *AJ*, *158*, 3, 109.
Hsu D. C. et al., 2020 *MNRAS*, *498*, 2, 2249.
Huber D. et al., 2014 *ApJS*, *211*, 1, 2.
Huber D. et al., 2017 *ApJ*, *844*, 2, 102.
Ida S. and Lin D. N. C., 2004 *ApJ*, *616*, 1, 567.
Jenkins J. M. and Dunnuck J., 2011 *Society of Photo-Optical Instrumentation Engineers (SPIE) Conference Series*, vol. 8146 of *Society of Photo-Optical Instrumentation Engineers (SPIE) Conference Series*, p. 814602.
Jenkins J. M. et al., 2010 *ApJL*, *713*, 2, L120.
Jenkins J. M. et al., 2015 *AJ*, *150*, 2, 56.
Johnson J. A. et al., 2010 *PASP*, *122*, 894, 905.
Jontof-Hutter D. et al., 2014 *ApJ*, *785*, 1, 15.
Jontof-Hutter D. et al., 2021 *AJ*, *161*, 5, 246.
Kawahara H. and Masuda K., 2019 *AJ*, *157*, 6, 218.
Kite E. S. and Schaefer L., 2021 *ApJL*, *909*, 2, L22.
Kite E. S. et al., 2019 *ApJL*, *887*, 2, L33.
Koch D. G. et al., 2010 *ApJL*, *713*, 2, L79.
Kopparapu R. K., 2013 *ApJL*, *767*, 1, L8.

Kunimoto M. and Matthews J. M., 2020 *AJ*, *159*, 6, 248.
Lambrechts M. et al., 2019 *A&A*, *627*, A83.
Latham D. W. et al., 2010 *ApJL*, *713*, 2, L140.
Latham D. W. et al., 2011 *ApJL*, *732*, 2, L24.
Lee E. J. and Chiang E., 2017 *ApJ*, *842*, 1, 40.
Lee E. J. and Connors N. J., 2021 *ApJ*, *908*, 1, 32.
Lillo-Box J. et al., 2021 *A&A*, *654*, A9.
Lindegren L. et al., 2018 *A&A*, *616*, A2.
Lissauer J. J., 1993 *ARA&A*, *31*, 129.
Lissauer J. J. and de Pater I., 2019 *Fundamental Planetary Science, Updated Edition*, Cambridge, UK: Cambridge University Press.
Lissauer J. J. and Eisberg J., 2018 *NewAR*, *83*, 1.
Lissauer J. J. et al., 2011a *Nature*, *470*, 53.
Lissauer J. J. et al., 2011b *ApJS*, *197*, 1, 8.
Lissauer J. J. et al., 2012 *ApJ*, *750*, 2, 112.
Lissauer J. J. et al., 2013 *ApJ*, *770*, 131.
Lissauer J. J. et al., 2014a *Nature*, *513*, 7518, 336.
Lissauer J. J. et al., 2014b *ApJ*, *784*, 1, 44.
Lissauer J. J. et al., 2023 *PSJ*, *in press*.
Lopez E. D. and Fortney J. J., 2013 *ApJ*, *776*, 1, 2.
Lopez E. D. and Fortney J. J., 2014 *ApJ*, *792*, 1, 1.
Lopez E. D. and Rice K., 2018 *MNRAS*, *479*, 4, 5303.
Loyd R. O. P. et al., 2020 *ApJ*, *890*, 1, 23.
Lundkvist M. S. et al., 2018 *Handbook of Exoplanets* (H. J. Deeg and J. A. Belmonte), p. 177.
MacDonald M. G. et al., 2021 *AJ*, *162*, 3, 114.
Marcy G. et al., 2005 *Progress of Theoretical Physics Supplement*, *158*, 24.
Marcy G. W. et al., 2014 *ApJS*, *210*, 2, 20.
Martinez C. F. et al., 2019 *ApJ*, *875*, 1, 29.
Masuda K., 2014 *ApJ*, *783*, 53.
Mathur S. et al., 2017 *ApJS*, *229*, 2, 30.
Mayor M. and Queloz D., 1995 *Nature*, *378*, 6555, 355.
Mayor M. et al., 2011 *arXiv e-prints*, arXiv:1109.2497.
Mazeh T. et al., 2012 *A&A*, *541*, A56.
Millholland S. and Laughlin G., 2017 *AJ*, *154*, 3, 83.
Millholland S. C. et al., 2021 *AJ*, *162*, 4, 166.
Mills S. M. and Fabrycky D. C., 2017 *ApJL*, *838*, 1, L11.
Mills S. M. et al., 2016 *Nature*, *533*, 509.
Moe M. et al., 2019 *ApJ*, *875*, 1, 61.
Moorhead A. V. et al., 2011 *ApJS*, *197*, 1, 1.
Mordasini C. et al., 2009 *A&A*, *501*, 3, 1161.
Morton T. D. and Johnson J. A., 2011 *ApJ*, *738*, 2, 170.
Morton T. D. et al., 2016 *ApJ*, *822*, 2, 86.
Muirhead P. S. et al., 2012 *ApJ*, *747*, 2, 144.
Mulders G. D. et al., 2015 *ApJ*, *798*, 2, 112.
Mulders G. D. et al., 2016 *AJ*, *152*, 6, 187.
Mulders G. D. et al., 2018 *AJ*, *156*, 1, 24.
Mulders G. D. et al., 2019 *ApJ*, *887*, 2, 157.
Mulders G. D. et al., 2021 *ApJL*, *920*, 1, L1.
Mullally F. et al., 2015 *ApJS*, *217*, 2, 31.
Nascimbeni V. et al., 2022 *A&A*, *658*, A31.
Nesvorný D., 2019 *NewAR*, *84*, 101507.
Nesvorný D. et al., 2013 *ApJ*, *777*, 3.
Newton I., 1687 *Philosophiae Naturalis Principia Mathematica. Auctore Js. Newton*.
Orosz J. A. et al., 2019 *AJ*, *157*, 5, 174.
Owen J. E. and Campos Estrada B., 2020 *MNRAS*, *491*, 4, 5287.
Owen J. E. and Wu Y., 2013 *ApJ*, *775*, 2, 105.
Owen J. E. and Wu Y., 2017 *ApJ*, *847*, 29.
Pascucci I. et al., 2019 *ApJL*, *883*, 1, L15.

Petigura E. A. et al., 2013a *ApJ*, *770*, 1, 69.
Petigura E. A. et al., 2013b *Proceedings of the National Academy of Science*, *110*, 48, 19273.
Petigura E. A. et al., 2017 *AJ*, *154*, 3, 107.
Petigura E. A. et al., 2018 *AJ*, *155*, 2, 89.
Pollack J. B. et al., 1996 *Icarus*, *124*, 1, 62.
Quintana E. V. et al., 2010 *Software and Cyberinfrastructure for Astronomy*, vol. 7740 of *Society of Photo-Optical Instrumentation Engineers (SPIE) Conference Series* (N. M. Radziwill and A. Bridger), p. 77401X.
Quintana E. V. et al., 2014 *Science*, *344*, 6181, 277.
Ragozzine D. and Holman M. J., 2018 *NewAR*, *83*, 5.
Rappaport S. et al., 2012 *ApJ*, *752*, 1, 1.
Rappaport S. et al., 2016 *MNRAS*, *458*, 4, 3904.
Rauer H. et al., 2014 *Experimental Astronomy*, *38*, 1-2, 249.
Raymond S. N. et al., 2007 *ApJ*, *669*, 1, 606.
Rizzuto A. C. et al., 2017 *AJ*, *154*, 6, 224.
Rogers J. G. and Owen J. E., 2021 *MNRAS*, *503*, 1, 1526.
Rogers J. G. et al., 2021 *MNRAS*, *508*, 4, 5886.
Rogers L. A., 2015 *ApJ*, *801*, 1, 41.
Rogers L. A. et al., 2011 *ApJ*, *738*, 1, 59.
Rowe J. F. et al., 2014 *ApJ*, *784*, 1, 45.
Rowe J. F. et al., 2015 *ApJS*, *217*, 1, 16.
Sanchis-Ojeda R. et al., 2013 *ApJ*, *774*, 1, 54.
Sanchis-Ojeda R. et al., 2014 *ApJ*, *787*, 1, 47.
Santerne A. et al., 2012 *A&A*, *545*, A76.
Santerne A. et al., 2016 *A&A*, *587*, A64.
Santos N. C. et al., 2004 *A&A*, *426*, L19.
Shallue C. J. and Vanderburg A., 2018 *AJ*, *155*, 2, 94.
Sheets H. A. and Deming D., 2014 *ApJ*, *794*, 2, 133.
Sheets H. A. and Deming D., 2017 *AJ*, *154*, 4, 160.
Smith J. C. et al., 2012 *PASP*, *124*, 919, 1000.
Sotin C. et al., 2007 *Icarus*, *191*, 1, 337.
Steffen J. H., 2016 *MNRAS*, *457*, 4, 4384.
Steffen J. H. and Agol E., 2005 *MNRAS*, *364*, 1, L96.
Steffen J. H. and Lissauer J. J., 2018 *NewAR*, *83*, 49.
Steffen J. H. et al., 2012a *Proceedings of the National Academy of Science*, *109*, 21, 7982.
Steffen J. H. et al., 2012b *MNRAS*, *421*, 2342.
Steffen J. H. et al., 2013 *MNRAS*, *428*, 1077.
Stumpe M. C. et al., 2012 *PASP*, *124*, 919, 985.
Stumpe M. C. et al., 2014 *PASP*, *126*, 935, 100.
Tenenbaum P. et al., 2010 *Software and Cyberinfrastructure for Astronomy*, vol. 7740 of *Society of Photo-Optical Instrumentation Engineers (SPIE) Conference Series* (N. M. Radziwill and A. Bridger), p. 77400J.
Tenenbaum P. et al., 2012 *ApJS*, *199*, 1, 24.
Thompson S. E. et al., 2016 Kepler Data Release 25 Notes, Kepler Science Document KSCI-19065-002.
Thompson S. E. et al., 2018 *ApJS*, *235*, 2, 38.
Torres G. and Fressin F., 2018 *NewAR*, *83*, 12.
Torres G. et al., 2011 *ApJ*, *727*, 24.
Tremaine S. and Dong S., 2012 *AJ*, *143*, 4, 94.
Twicken J. D. et al., 2010 *Software and Cyberinfrastructure for Astronomy*, vol. 7740 of *Society of Photo-Optical Instrumentation Engineers (SPIE) Conference Series* (N. M. Radziwill and A. Bridger), p. 774023.
Valencia D. et al., 2007 *ApJ*, *665*, 2, 1413.
Van Eylen V. et al., 2018 *MNRAS*, *479*, 4786.
Wang J. and Fischer D. A., 2015 *AJ*, *149*, 1, 14.
Weiss L. M. et al., 2018 *AJ*, *155*, 1, 48.
Williams J. P. and Cieza L. A., 2011 *ARA&A*, *49*, 1, 67.

Winn J. N. and Fabrycky D. C., 2015 *ARA&A*, *53*, 409.
Winn J. N. et al., 2017 *AJ*, *154*, 6, 270.
Winn J. N. et al., 2018 *NewAR*, *83*, 37.
Wolniewicz L. M. et al., 2021 *AJ*, *161*, 5, 231.
Wright J. T. et al., 2012 *ApJ*, *753*, 2, 160.
Xie J.-W., 2013 *ApJS*, *208*, 22.
Xie J.-W. et al., 2016 *Proceedings of the National Academy of Science*, *113*, 41, 11431.
Youdin A. N., 2011 *ApJ*, *742*, 1, 38.
Zhang Z. et al., 2023 *AJ*, *165*, 2, 73.
Zhu W., 2019 *ApJ*, *873*, 1, 8.
Zhu W. et al., 2018 *ApJ*, *860*, 2, 101.
Zink J. K. and Hansen B. M. S., 2019 *MNRAS*, *487*, 1, 246.

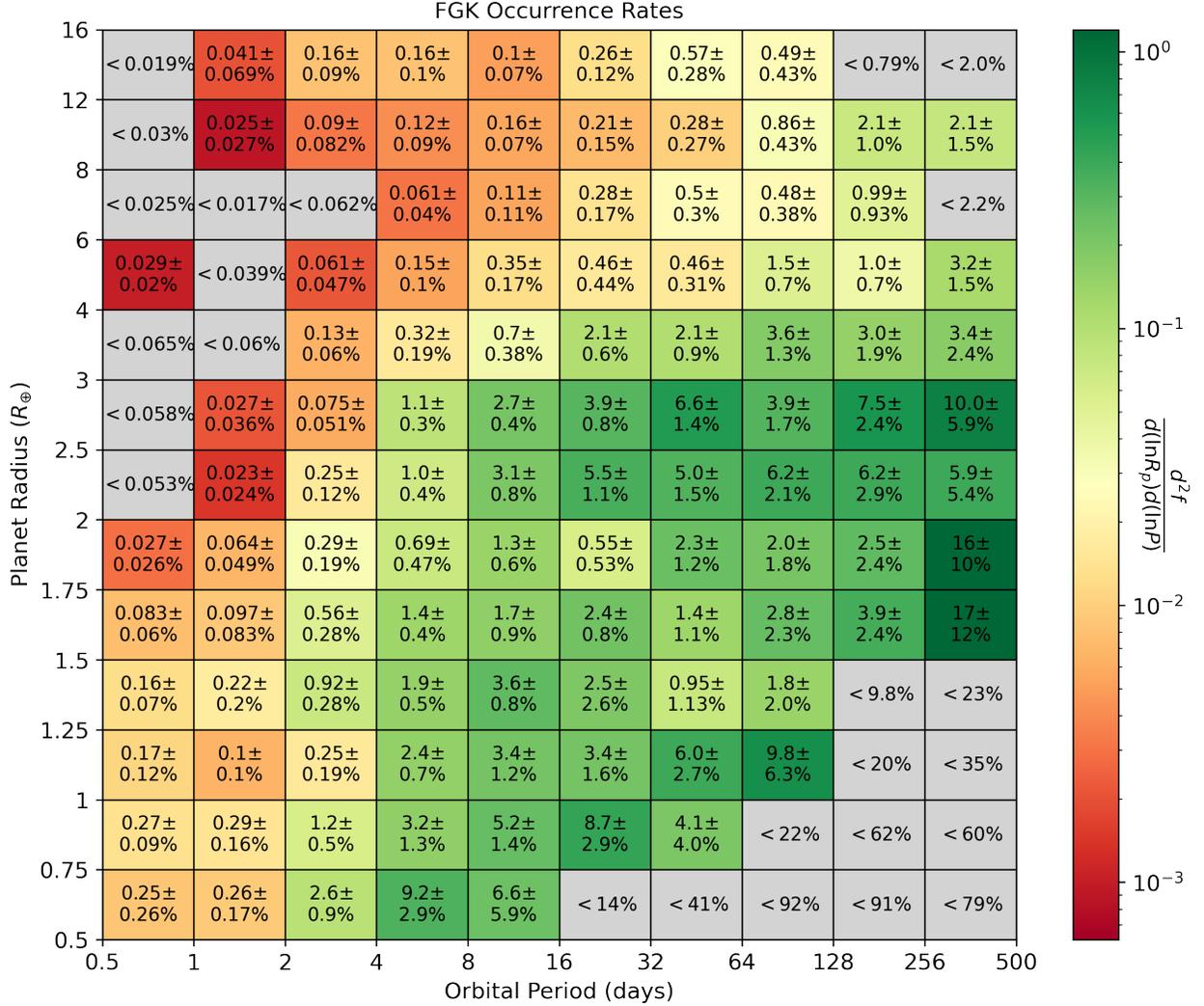

Fig. 2.— Occurrence rates expressed as the average number of planets within each of the planetary radius and orbital period bins per FGK-type star. Note that the numbers are given as percentage, and if one drops the % sign, they represent number of planets per 100 stars. The values shown were computed from planets detected in the DR25 catalog, with corrections made to compensate for detection incompleteness and the geometrical factors. Colors indicate how common planets are per logarithmic bin in $R_p$ and $P$, with dark green representing the highest values and dark red the lowest. For example, the bin with the highest estimated occurrence rate, planets with 1.5 $R_\oplus < R_p <$ 1.75 $R_\oplus$ and 256 days $< P <$ 500 days, has $17 \pm 12$ planets per 100 stars. Gray indicates portions of parameter space with few detected planets, for which only upper limits on occurrence rates are known. Data from *Hsu et al.* (2019). Graphic courtesy Michelle Kunimoto, Danley Hsu & Eric Ford.

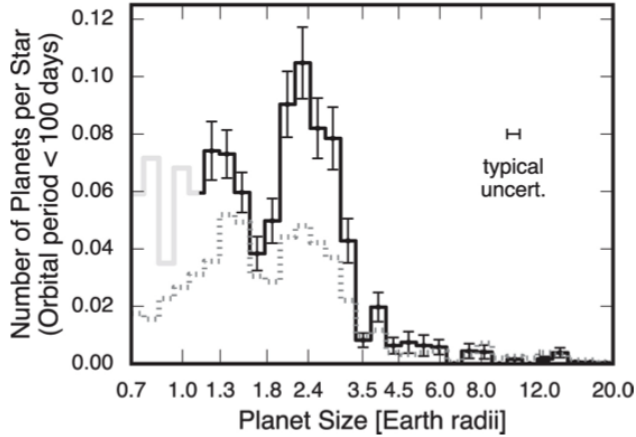

Fig. 3.— The solid line shows the estimated number of planets per sunlike star with orbital periods $P < 100$ days as a function of planet size. The dotted line shows the size distribution of detected planets, before completeness corrections scaled for visual comparison. The solid line is shown in light gray for planets of roughly Earth size and smaller to emphasize the large uncertainties in correcting for undetected planets in this size range. From *Fulton and Petigura* (2018).

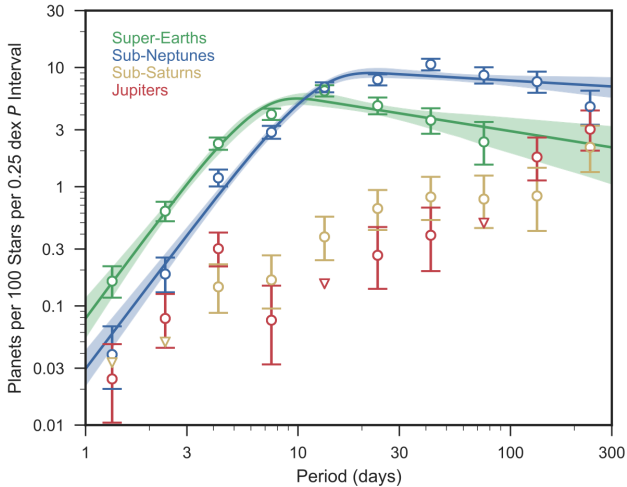

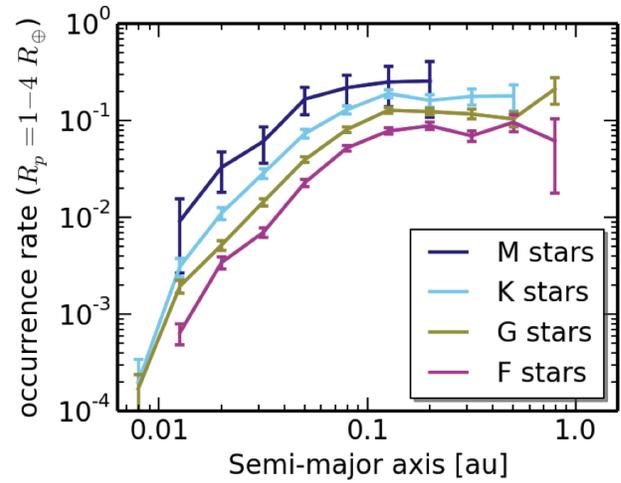

Fig. 5.— Occurrence rates of planets with radii $1\,R_\oplus < R_p < 4\,R_\oplus$ as functions of orbital semimajor axis for dwarf stars of different spectral types. Smaller stars are seen to host more planets in this size range at all orbital distances for which adequate data are available. From *Mulders et al.* (2015).

Fig. 4.— Planet occurrence as a function of orbital period for various ranges of planetary radii. Points show the number of planets in their size bin per 100 stars per factor of $10^{1/4} \approx 1.78$ in the period. Planets smaller than Neptune are well-fit by broken power laws, with nearly constant numbers per unit $\log P$ at $P > 7$ days for super-Earths and at $P > 12$ days for sub-neptunes and steep decreases for shorter periods. In contrast, the abundances of larger planets tend to increase for longer periods, apart from a moderately large bump for hot jupiters with $P \sim 4$ days and possibly a smaller bump for hot sub-saturns with similar periods. From *Petigura et al.* (2018).

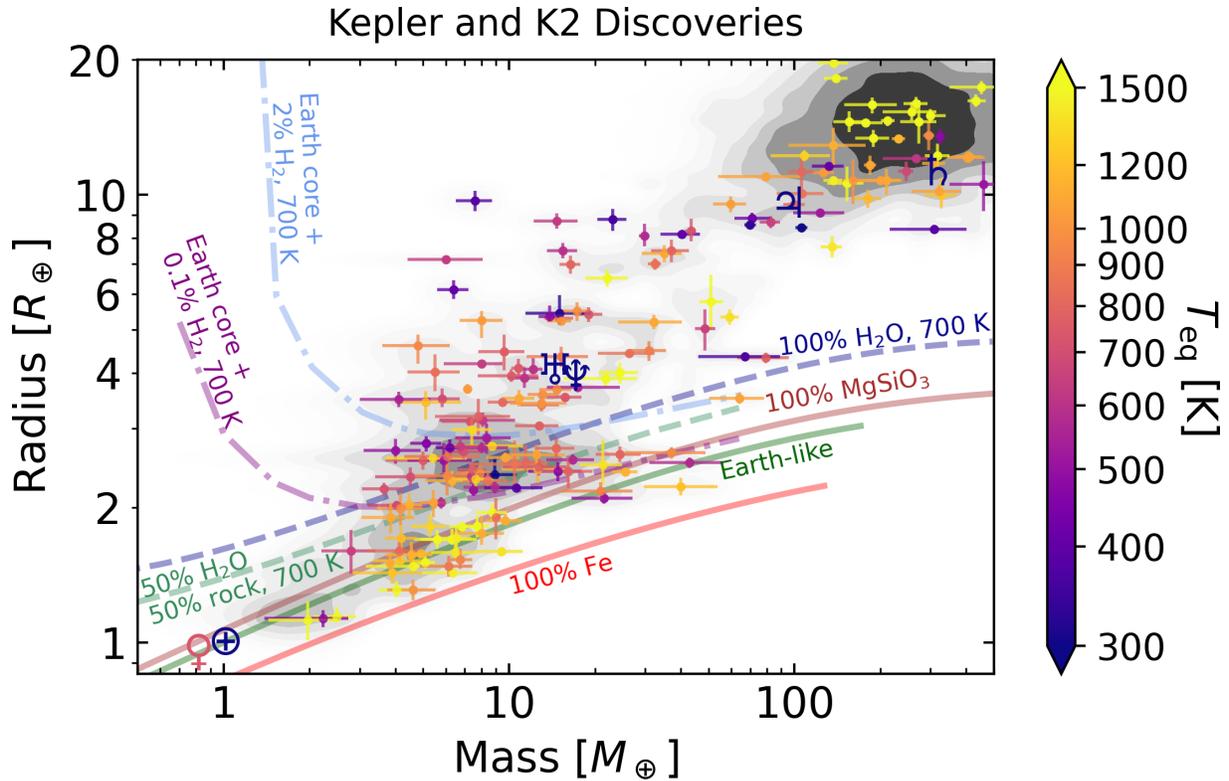

Fig. 6.— Mass-radius diagram for exoplanets discovered by *Kepler* and *K2* with precisely-measured masses. The background grayscale shading map includes all transiting planets with precise masses. Colored lines indicate theoretical models with the compositions stated. Solid curves are used for planets entirely composed of refractory materials, whereas broken curve represent planets with volatile components (water or $H_2$). Here "Earth core +" means that the interior of the planet has the same composition as the bulk Earth (including the mantle), and it is surrounded by a hydrogen envelope that contributes the fraction of the planet's mass specified. Note that giant planets dominate the overall data, but small and mid-sized planets make up the majority of *Kepler* and *K2* discoveries. Courtesy Joseph Murphy.

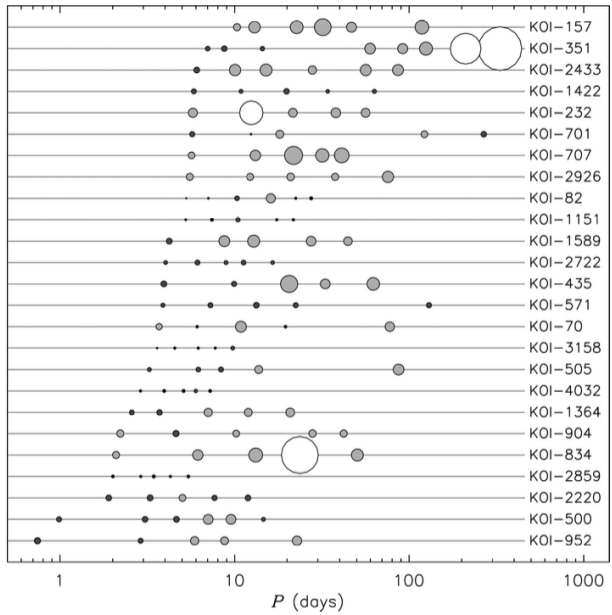

Fig. 7.— *Kepler*'s candidate multi-planet systems containing five or more planets. Each line corresponds to one system, as labeled on the right side. Ordering is by the innermost orbital period. Planet radii are to scale relative to one another, with small planets ($R_p < 1.7$ R$_\oplus$) represented by black circles, planets with radii $1.7$ R$_\oplus < R_p < 6$ R$_\oplus$ shown in gray and large planets ($R_p > 6$ R$_\oplus$) represented by open circles. Plot from *Lissauer and de Pater* (2019); courtesy Daniel Fabrycky.

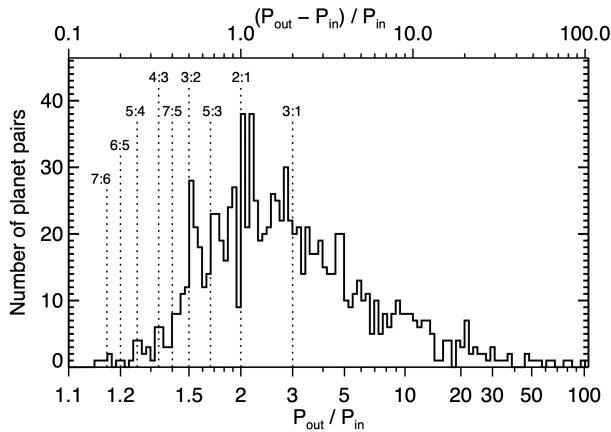

Fig. 8.— Period ratios of pairs of planets in *Kepler*'s candidate multi-planet systems. The top axis shows the fractional period difference, and the histogram bin widths represent 1/40 of the logarithm of this quantity. The bottom axis indicates the period ratio. From *Winn and Fabrycky* (2015).

23